\documentclass[%
 reprint,
 amsmath,amssymb,
 aps,
]{revtex4-1}
\usepackage{appendix}
\usepackage{graphicx}% Include figure files
\usepackage{dcolumn}% Align table columns on decimal point
\usepackage{bm}% bold math
\usepackage{tabularx}
\usepackage{setspace}
\usepackage{xcolor}
\usepackage[portrait, margin=0.5in]{geometry}

\begin{document}
\renewcommand{\arraystretch}{1.5}
% \renewcommand{\baselinestretch}{1.5}
%\preprint{APS/123-QED}

\title{Phonon eigenfunctions of inhomogeneous lattices: Can you hear the shape of a cone?}% Force line breaks with \\

\author{Grace H. Zhang}
%\email{ghzhang@g.harvard.edu}
\affiliation{Department of Physics, Harvard University, Cambridge, MA 02138, USA.}
\author{David R. Nelson}
\affiliation{Department of Physics, Harvard University, Cambridge, MA 02138, USA.}%

\date{\today}
% \pagenumbering{gobble}

\begin{abstract}
We study the phonon modes of interacting particles on the surface of a truncated cone resting on a plane subject to gravity, inspired by recent colloidal experiments. We derive the ground state configuration of the particles under gravitational pressure in the small cone angle limit, and find an inhomogeneous triangular lattice with spatially varying density but robust local order. The inhomogeneity has striking effects on the normal modes such that an important feature of the cone geometry, namely its apex angle, can be extracted from the lattice excitations. The shape of the cone leads to energy crossings at long wavelengths and frequency-dependent quasi-localization at short wavelengths. We analytically derive the localization domain boundaries of the phonons in the limit of small cone angle and check our results with numerical results for eigenfunctions. 

\end{abstract}

\pacs{Valid PACS appear here}% PACS, the Physics and Astronomy
%  \tableofcontents             
% \clearpage
\maketitle
\normalsize
  
\section{\label{sec:intro}Introduction}

“Can you hear the shape of a drum?” is a classic problem formulated by Mark Kac: Can we determine the geometry of a two-dimensional membrane from the eigenvalues of the Laplace equation~\cite{kac1966can}? 
This seminal work led to a proliferation of studies on extracting geometrical information from a system's normal modes. For example, Ref.~\cite{mitra1992diffusion} examines how the geometry of the pore-grain interface in a porous media can be recovered from the eigenvalues and eigenfunctions of the diffusion propagator. 
These works have focused on the membrane as a continuous medium. 
In this paper, we study the effect of a discrete periodic lattice structure with a slowly varying lattice constant on problems of this category. We focus on truncated cones, and as in particular whether the fundamental shape parameter of the cone angle can be inferred from the normal modes. 

When interacting particles ordered in a two-dimensional plane are under inhomogeneous strains (e.g. due to a gravitational pressure), they can in general remove the extra energy cost, a form of geometrical frustration, in at least two distinct ways: by forming topological defects in an otherwise homogeneous lattice, or by maintaining a defect-free lattice by curving into the third dimension \cite{menezes2019self,silva2020formation,ansell2020transitions}. 
We study here the latter scenario. As we shall see, the overall shape of the curved membrane is coupled with lattice spacings that can generally be non-uniform, converting the lattice into an inhomogeneous crystal. 

Inhomogeneous crystals are collections of particles with nonuniform equilibrium densities but robust well-ordered local structures. 
Experimentally, such lattices have been observed in plasmas under  gravitational or flow fields~\cite{thomas1994plasma,mughal2007topological,thomas1996melting}, foams between curved surfaces~\cite{drenckhan2004demonstration}, ions in laser traps~\cite{mielenz2013trapping}, and, most recently, colloidal particles confined by an applied electric field~\cite{soni2018emergent}. However, a complete understanding of inhomogeneous crystals remains elusive, because the inhomogeneity of the lattice seems incompatible with ideas from condensed matter physics that typically assume perfect periodicity. 

In this paper, we study a two-dimensional (2d) inhomogeneous crystal, consisting of interacting particles that deform from a cylindrical shell into a truncated cone under gravitational hydrostatic pressure. 
In the spirit of inverse problems, can we infer information on the shape of this cone (such as the inclination angle) from its normal modes (in this case, the in-plane phonons of an inhomogeneous lattice)? 
We answer in the affirmative. Slow adiabatic changes in the lattice spacing, controlled by the cone angle, manifest at long wavelengths as a reordering of low energy eigenmodes, and at short wavelengths as quasi-localization due to a spatially varying local band edge, the maximum energy (or frequency) beyond which lattice phonons cannot propagate. Our results could be checked experimentally, for example, on systems of overdamped density-mismatched colloidal particles adsorbed onto conical surfaces~\cite{sun2020colloidal,rehman2019self}. 

The paper is organized as follows. In Sec.~\ref{sec:pileup}, as a simplified warm up problem, we illustrate inhomogeneity-induced quasi-localization of lattice excitations using a one-dimensional (1d) model of dislocation pileups~\cite{hirth1983theory,zhang2020pileups}, which exhibits the same phenomenon as the 2d conic sheet studied in the main body of the paper. In Sec.~\ref{sec:shape}, we argue that a truncated cone with a slowly varying lattice constant is the ground state of interacting particles with cylindrical boundary conditions under gravitational pressure. In Sec.~\ref{sec:phonons}, we study the effect of inhomogeneity on the in-plane phonons. We find that low frequency phonons exhibit interesting energy crossings as a function of the cone angle (Sec.~\ref{sec:low_phonons}) and high frequency phonons are localized inside spatial domains whose sizes depend on the cone angle. We derive the localization domain boundaries as a function of the phonon frequency and the angle of the cone, using two complementary methods: We recover this boundary (1) in momentum space by identifying the spatially-varying band edge (Sec.~\ref{sec:high_phonons_BE}), and (2) in real space by a mapping to Schrodinger's equation and applying an inverted Wentzel–Kramers–Brillouin (WKB) analysis (or the Liouville-Green method, also due to Jeffreys)~\cite{wentzel1926verallgemeinerung,kramers1926wellenmechanik,brillouin1926mecanique,jeffreys1925certain,green1838motion} (Sec.~\ref{sec:high_phonons_WKB}). 

\section{One-dimensional example: dislocation pileups \label{sec:pileup}}

We expect that the localization effect we find for inhomogeneous cones (Sections \ref{sec:shape} and \ref{sec:phonons}) occurs more generally in inhomogeneous lattices. Here, we illustrate this phenomenon in a relatively simple one-dimensional inhomogeneous lattice with long range interactions and only longitudinal phonons, inspired by the physics of dislocation pileups~\cite{hirth1983theory,zhang2020pileups}. Specifically, we show that the normal modes of 1d dislocation pileups, a type of defect assembly embedded in two-dimensional crystals~\cite{hirth1983theory}, exhibit quasi-localization as a function of the phonon frequency, due to a position-dependent band edge. 
% This section is mainly intended as a pedagogical introduction to the concepts discussed later in the paper; 
Readers are referred to Appendix \ref{app:pileup} for details and may skip ahead to Sec.~\ref{sec:shape} without loss of continuity. 

Dislocation pileups exemplify an intriguing class of 1d inhomogeneous lattices, whose constituents are not particles but point-like edge dislocation defects in a two-dimensional host crystal~\cite{hirth1983theory}. The dislocations, once emitted in response to a stress field, all have the same Burgers vector topological charge and interact via a long-ranged logarithmic potential, with an average dislocation density determined by the form of the external shear stress~\cite{zhang2020pileups}. 

\begin{figure}[htb]
\centering
\includegraphics[width=0.95\columnwidth]{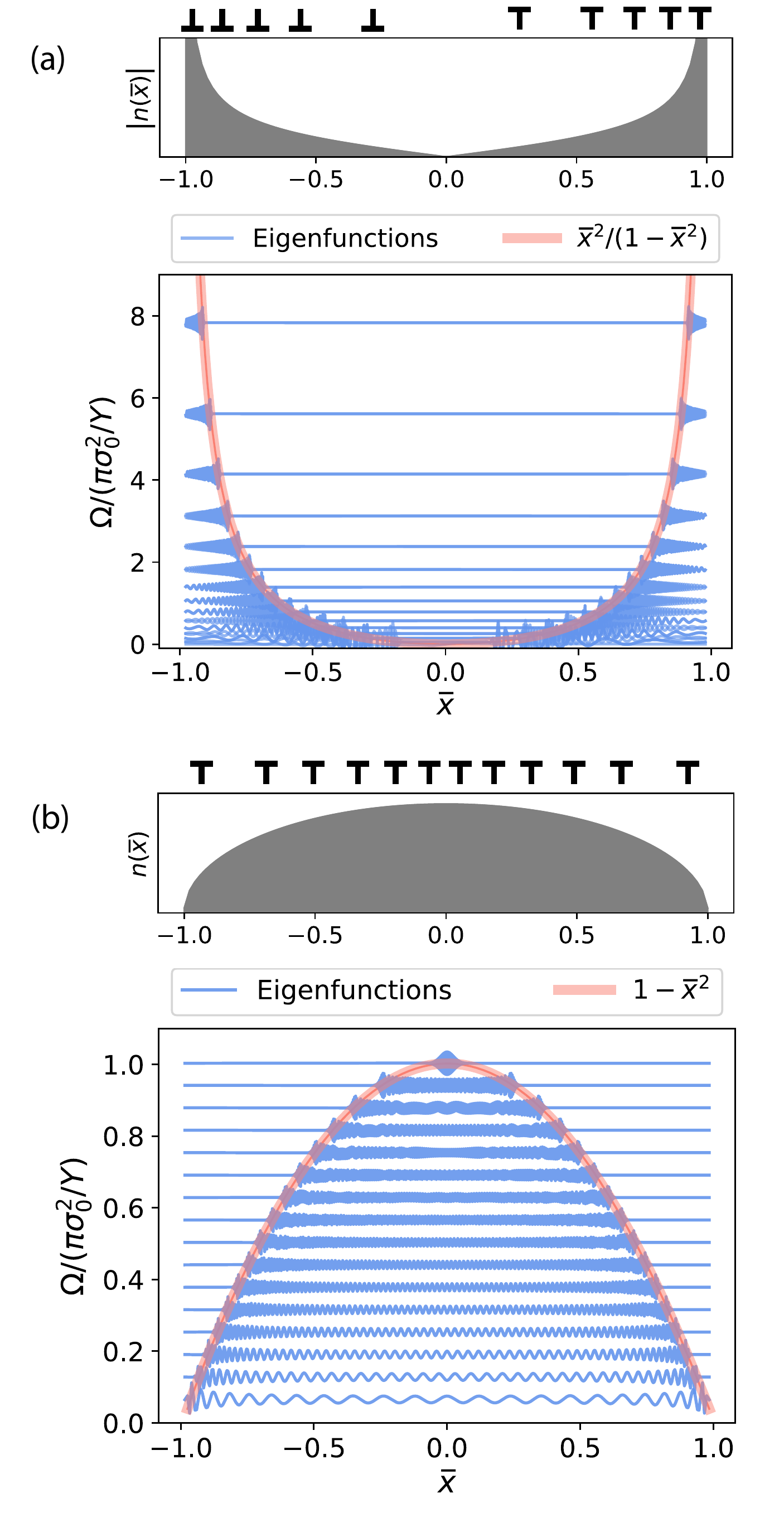}%
\caption{\setstretch{1.} The localization boundaries in one dimension predicted by Eq.~(\ref{eq:tp_all_springs}) (red lines) and the numerical phonon eigenfunctions (blue) for two types of pileups with different average dislocation densities $n(\bar x)$, where $\bar x \equiv x/(L/2)$, plotted in gray and illustrated by schematics of dislocations. The eigenfunctions are stacked vertically according to their rescaled eigenenergies $\Omega$. As predicted by Eq.~(\ref{eq:tp_all_springs}), the localization boundaries are proportional to $n(x)^2$: the eigenfunctions localize towards the densest parts of the dislocation lattice (the edges in (a) and the center in (b)) as $\Omega$ increases. \label{fig:Vfull}}
\end{figure}

Fig.~\ref{fig:Vfull} shows the eigenfunctions for two types of pileups, resulting from applied stress fields in the two-dimensional host lattice that are uniform (Fig.~\ref{fig:Vfull}a) and linearly varying (Fig.~\ref{fig:Vfull}b) in space. The eigenfunctions are stacked vertically according to their eigenenergies $\Omega$. For both types of pileup, the eigenfunctions become more localized as $\Omega$ increases, with a localization domain concentrated within the densest parts of the pileup as the eigenmode frequency increases. 

We find the localization domain boundaries of the longitudinal pileup phonon modes by calculating the band edge of the phonon dispersion for a uniform pileup (see Appendix \ref{app:pileup} or Ref.~\cite{zhang2020pileups}), given by the simple condition,
\begin{equation} \label{eq:pileup_BE}
    \Omega_\text{edge} = \frac{Yb^2 \pi}{16 D^2},
\end{equation}
where $D$ is the average dislocation spacing of the pileup, $Y$ is the Young's modulus of the 2d host crystal, and $b$ is the magnitude of the dislocation Burger's vector. 
% Now, let $k_\alpha = \frac{2 \pi}{\lambda_\alpha}$ be the momentum associated with the $\alpha$-th eigenstate, where $\lambda_\alpha$ is the associated wavelength. In an inhomogeneous lattice, the lattice constant $D(x)$ varies as a function of space $x$. When $\frac{\lambda_\alpha}{2}> D(x)$, 
Eigenmodes with frequencies above this threshold $\Omega > \Omega_\text{edge}$ enter the band gap and decay exponentially, since the dislocation lattice cannot resolve and propagate the phonon, instead generating strong back scattering (Bragg diffraction in one dimension). When the dislocation lattice spacing is space-dependent $D = D(x)$, as in the case for the pileups shown in Fig.~\ref{fig:Vfull}, the local band edge becomes slowly varying in space, with $\Omega = \Omega(x)$. On using Eq.~(\ref{eq:pileup_BE}), the localization domain $[x^-, x^+]$ and energy $\Omega$ of the band edge satisfies,
\begin{eqnarray} 
\Omega(x^\pm) &=&  \frac{Y b^2 \pi }{16}  n(x^\pm)^2. \label{eq:tp_all_springs}
\end{eqnarray}
where $n(x)$ is the average dislocation charge density, related to the dislocation lattice spacing via $D(x) = |n(x)|^{-1}$, given by 
\begin{eqnarray}
n_{\mathrm{D}}(x) &=& \frac{4 \sigma_{0}}{Y b} \frac{x}{\sqrt{\left(\frac{L}{2}\right)^{2}-x^{2}}} \\
n_{\mathrm{SC}}(x) &=&\frac{4 \sigma_{0}}{Y b} \sqrt{1-\left(\frac{x}{L / 2}\right)^{2}}
\end{eqnarray}
for the eigenfunctions shown in Fig.~\ref{fig:Vfull}, and $Y$ is the Young's modulus of the 2d crystal, $b$ is the magnitude of the dislocation Burgers vector, and $\sigma_0$ is the strength of the applied shear stress. 
Note that the domain of localization tracks the square of the dislocation density profile $\sim n(x)^2$.
Eq.~(\ref{eq:tp_all_springs}) is plotted in red Fig.~\ref{fig:Vfull} and shows good agreement with the localization domains of our numerical eigenfunctions for the two dislocation densities shown. 

Nonlinear interactions between the dislocation degrees of freedom is essential for the quasi-localization phenomena. In contrast, a chain of balls and springs with constant stiffness on an inclined surface, which also assumes an inhomogeneous configuration, does \textit{not} exhibit quasi-localization because the interaction strengths embodied in the spring constants are independent of the inhomogeneous lattice spacing (see Appendix \ref{app:1d} for details).

\section{Cone Shape \label{sec:shape}}

In this section, we introduce a particularly simple two-dimensional microscopic model with an inhomogeneous lattice constant: particles initially within a 2d cylindrical sheet interacting via a Lennard-Jones (LJ) pair potential, which then deforms under gravitational pressure (Sec.~\ref{sec:micro} below). 
Unlike a sheet in the flat plane with gravity pointing downwards in the plane of the sheet (see Appendix~\ref{app:sheet}), a cylindrical shell subjected to a weak gravitational field parallel to the cylinder axis relieves its frustration by transforming into a truncated cone with an inhomogeneous lattice spacing (see Sec.~\ref{sec:elas}). In Sec.~\ref{sec:conn}, we map our discrete particle model onto continuum elasticity theory to derive the cone angle as a function of the microscopic parameters and the spatially varying lattice spacing as a function of the cone surface coordinates. (In Sec.~\ref{sec:phonons}, we will show that information about the cone angle can also be inferred from the phonon spectrum.)

\subsection{Microscopic model \label{sec:micro}}

\begin{figure}[htb]
\centering
\includegraphics[width = 1\columnwidth]{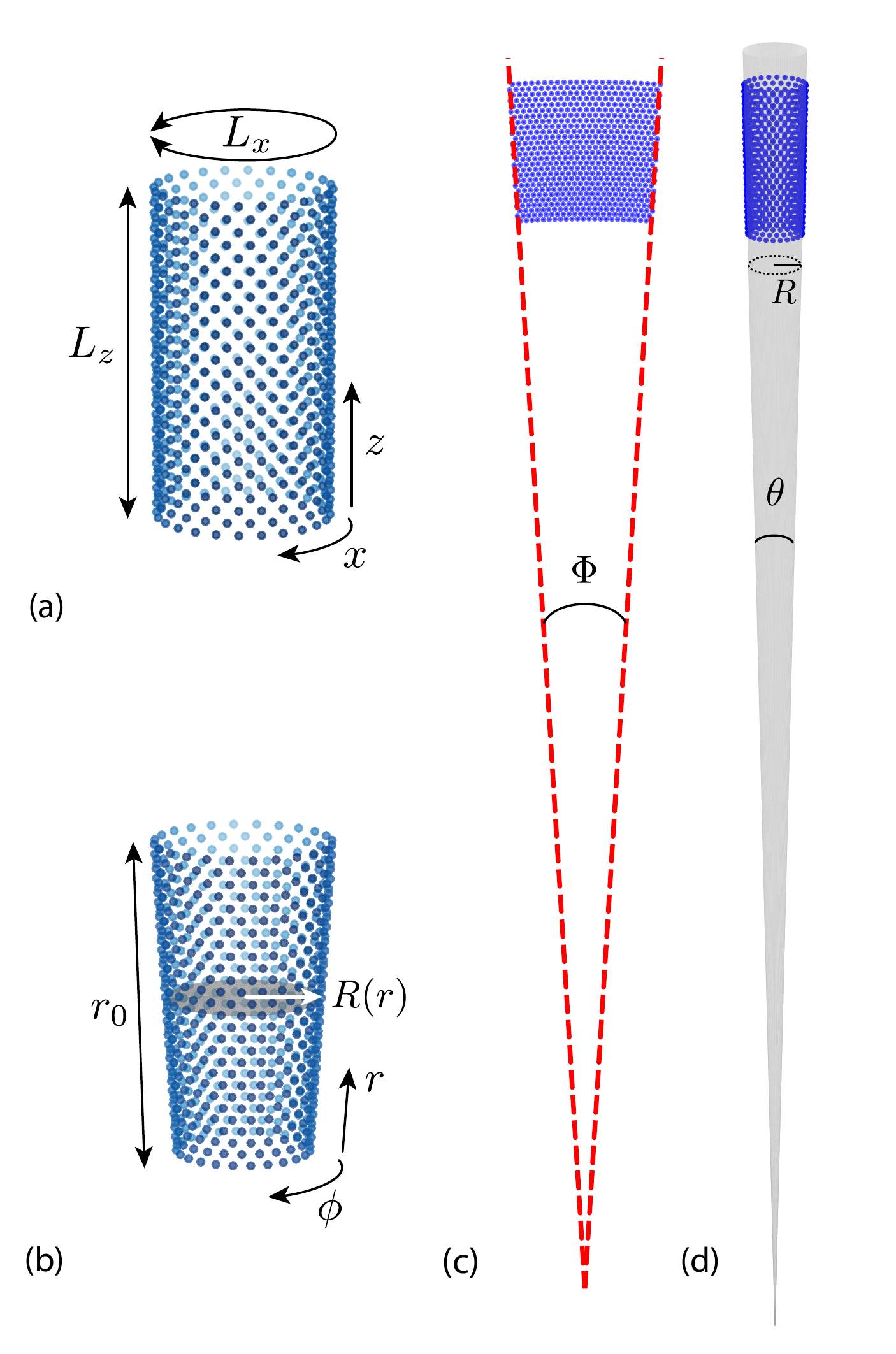}
\caption{(a): The ground state configuration of $N$ particles interacting via the Lennard-Jones potential in a cylindrical sheet is a triangular lattice with lattice constant $a_0$, height $L_z$, and rim circumference $L_x$. (b): Under gravitational pressure, the cylinder deforms into a truncated cone with length $r_0$. The coordinates of the cone are the azimuthal angle $\phi$, the longitudinal length along the surface $r$. The cross-sectional radius $R = R(r)$ increases as a function of $r$. (d)-(e): The sector angle $\Phi$ of the rolled out cone (d) is related to the apex angle $\theta$ (e) by $\theta = \Phi/\pi$. The numerical parameters used in this figure are $(L_z, L_x, a_0, N, \theta) = (13 \sqrt{3}, 24, 1, 648, 1.1^\circ)$. }
\label{fig:grav}
\end{figure}

Consider a 2d system of particles interacting via a pair potential $V_\text{int}$ and experiencing an external one-body potential $V_\text{ext}$:
\begin{eqnarray} \label{eq:H_micro}
H[ \{ \vec r_n \}] =  \sum_{n \neq m} V_\text{int} (\vec r_n, \vec r_m) + \sum_{n = 1}^N  V_\text{ext}(\vec r_n)
\end{eqnarray}
where $N$ is the total number of particles and $n,m$ are particle indices. For concreteness, we work with an isotropic interaction potential $V_\text{int} (\vec r, \vec r') = V_\text{int}(|\vec r - \vec r'|)$ of the Lennard-Jones form,
\begin{eqnarray} \label{eq:LJ}
V_{\mathrm{int}} (r) =\varepsilon\left[\left(\frac{a_0}{r}\right)^{12}-2\left(\frac{a_0}{r}\right)^{6}\right],
\end{eqnarray}
which has both a repulsive and an attractive component, with a potential minimum at $r = a_0$. 
(Upon setting $\sigma \equiv 2^{- 1/6} a_0 $, Eq.~(\ref{eq:LJ}) can be rewritten in the usual way, $V_{\mathrm{int}} (r) =4 \varepsilon\left[\left(\frac{\sigma}{r}\right)^{12}-\left(\frac{\sigma}{r}\right)^{6}\right]$.)
Without an external potential, particles in the ground state form a triangular lattice with a uniform lattice constant $a_0$ (see Fig.~\ref{fig:grav}a)~\cite{ashcroft1978solid}. However, as shown in the next section, the classical ground state can be distorted in an interesting way by a nonzero gravitational potential, 
\begin{eqnarray} \label{eq:V_ext}
V_\text{ext}(\vec r) = mg z,
\end{eqnarray}
where $m$ is the effective particle mass and $g$ is the gravitational acceleration. In colloidal experiments, $m$ can be changed by tuning the density mismatch between the colloid and the solvent. 
The pair potential in Eq.~\ref{eq:H_micro} is assumed to act in three dimensions, along a chord connecting atoms embedded in a cone or cylinder. In settings where the colloids are adsorbed onto a cylindrical or conical template, or the bending rigidity is very large, flexural phonons are frozen out. Henceforth, we ignore flexural phonons and focus only on the physics of in-plane fluctuations in Sec.~\ref{sec:phonons}. 
% The force balance condition at equilibrium is obtained by taking the spatial derivative of Eq.~\ref{eq:H_micro}:
% \begin{eqnarray} \label{eq:force_b}
% \frac{\partial V_\text{ext} (\vec R_n)}{\partial R_{n, i}} = \sum_{m} At \frac{ (\vec R_n - \vec R_m)_{i}}{|\vec R_n - \vec R_m|^{t+2}}
% \end{eqnarray}
% where

The microscopic model in Eq.~(\ref{eq:H_micro})-(\ref{eq:V_ext}) can be mapped at long wavelengths onto continuum elasticity theory. 
We use the latter tool to calculate the inhomogeneous equilibrium lattice positions under the gravitational potential in the next section, and elaborate on the details of this mapping in Sec.~\ref{sec:conn}.  

\subsection{Elasticity theory \label{sec:elas}}

We first examine a 2d cylindrical sheet with uniform elastic coupling constants using continuum elasticity theory and then calculate the equilibrium displacements under gravitational pressure. As detailed in Appendix~\ref{app:force}, these displacements hold for the lattice of interacting particles in Eq.~(\ref{eq:H_micro}) when the gravitational pressure is sufficiently weak such that,
\begin{eqnarray} \label{eq:small_reg}
\frac{\alpha_0}{2B}  \frac{L_z }{2}\ll 1,
\end{eqnarray}
where $B = \mu + \lambda$ is the bulk modulus of the unperturbed crystal. 
In this limit, the resulting configuration describes a truncated cone (Fig.~\ref{fig:grav}b), with the precise correspondence shown in Eqs.~(\ref{eq:az})-(\ref{eq:acone}) below. Thus, a cylinder whose circular base is free to slide on the flat surface on which it rests, collapses downward and inward under gravitational pressure, resulting in the cylinder base being smaller and denser than its less deformed upper end. 

A two-dimensional cylindrical sheet, with periodic boundary conditions imposed along $x$ and the axis of symmetry pointing along $z$, under a gravitational potential can be studied as for the similar problem of a 3d elastic solid subject to gravity, as discussed in Ref.~\cite{landau2012theory}. Under an isotropic pressure that increases with decreasing vertical coordinate $z$ (lower portions of the sheet bear more weight from above), the cylindrical sheet's free energy is given by
\begin{eqnarray} \label{eq:F_grav_cyl}
F = \frac{1}{2} \int d^2 r  \left( 2 \mu u_{ik}^2 + \lambda u_{ii}^2  + 2 \alpha_0 (L_z-z) u_{ii} \right),
\end{eqnarray}
where $\delta p = \alpha_0 (L-z)$ is the isotropic pressure, $u_{ij} = \frac{1}{2} (\frac{\partial u_i}{\partial r_j} + \frac{\partial u_j}{\partial r_i} ) $ is the strain tensor, $\mu$ and $\lambda$ are the first and second 2d Lam\'{e} coefficients. Here $\alpha_0$ is proportional to the gravitational constant $g$ controlling the hydrostatic pressure of, say, density-mismatched colloids in a solvent (see Eq.~(\ref{eq:alpha0_g}) below) and $\vec u(\vec r)$ is the displacement of the sheet from its  $\alpha_0 = 0$ equilibrium configuration at position $\vec r$.  
The non-deformed ($\alpha_0 = 0$) cylindrical sheet has a height of $L_z$ and a rim circumference of $L_x$. The natural boundary conditions for this problem are (1) anchoring the bottom edge of the sheet at the base, which requires that the vertical displacements $u_z$ vanish at $z = 0$, and (2) the stresses vanish at the upper edge $z = L_z$ of the truncated cone. 

In the absence of pressure ($\alpha_0 = 0$), the free energy in Eq.~(\ref{eq:F_grav_cyl}) is minimized when the equilibrium displacements $\bar u_i$ vanish and the strain tensor $\bar u_{ij} = 0$ for all $i,j$. 
We now calculate the equilibrium positions under nonzero $\alpha_0$ by finding the set of uniform strains $\bar u_{ij}$ that minimize Eq.~(\ref{eq:F_grav_cyl}). Since the gravitational term is isotropic and does not depend on $u_{zx}$, we immediately have $\bar u_{zx} = 0$. 
Upon setting the derivatives of the free energy $F$ with respect to $u_{zz}$ and $u_{xx}$ to zero and using the boundary conditions $\bar u_{zz}=\bar u_{xx}=0$ at $z = L_z$, we obtain the polar projection of the displacements as (see Fig.~\ref{fig:grav}c and Appendix \ref{app:sheet} for details),
\begin{eqnarray} 
\bar u_z &=& \frac{\alpha_0}{2 B} \left( \frac{z^2}{2} - L_z z - \frac{x^2}{2}\right) \label{eq:solx}\\
\bar u_x &=& \frac{\alpha_0}{2 B} (z - L_z) x. \label{eq:soly}
\end{eqnarray}
where $B = \mu + \lambda$ is the 2d bulk modulus. Upon assuming the gravitational pressure to be sufficiently weak (Eq.~(\ref{eq:small_reg})) and imposing periodic boundary conditions by wrapping the sheet around in the $x$ direction to form a cone, we obtain the axial and azimuthal displacements in terms of the surface coordinates as,
\begin{eqnarray} 
\bar u_r &=& \frac{\alpha_0}{2 B} \left( \frac{r^2}{2} - r_0 r\right) \label{eq:solr}\\
\bar u_\phi &=& \frac{\alpha_0}{2 B} \frac{L_x}{2 \pi} (r - r_0) \phi, \label{eq:solrad}
\end{eqnarray}
where $r$ is the longitudinal coordinate along the surface, $r_0$ is the longitudinal length of the truncated cone, $\phi$ is the azimuthal angle around the cone axis (see Fig.~\ref{fig:grav}b), and the zero displacement boundary condition at the bottom edge $\bar u_{r} = 0$ at $r = 0$ is now satisfied, in contrast to Eq.~(\ref{eq:solx}). 

% In a sense, the implicit Gaussian curvature at the tip of the cone cancels out the geometric frustration that led to the extra strain energy in flat space.  
Upon redefining $u_i \rightarrow u_i - \bar u_i$  in Eq.~(\ref{eq:F_grav_cyl}), where $\bar u_i$ is the equilibrium displacement of the cone relative to the cylinder, we can eliminate (gauge away) the pressure term in the free energy (see Appendix~\ref{sec:F0F} for details) and obtain, up to an additive constant,
\begin{eqnarray} \label{eq:F_elas}
F = \frac{1}{2} \int d^2 r  \left( 2 \mu u_{ik}^2 + \lambda u_{ii}^2  \right). 
\end{eqnarray}

We now calculate the apex angle $\theta$ of the cone (Fig.~\ref{fig:grav}d), as a function of the gravitational pressure coefficient $\alpha_0$ and elastic constants $\mu$ and $\lambda$, from the sector angle of an unfolded cone given by (see Fig.~\ref{fig:grav}c),
\begin{eqnarray} \label{eq:Phi_def}
\tan \frac{\Phi}{2} = \frac{\Delta x}{\Delta z},
\end{eqnarray}
where,
\begin{equation}
\begin{aligned} \label{eq:sector_calc}
\Delta x &= \bar u_x \left( x = \frac{L_x}{2}, z = L_z \right ) - \bar u_x \left( x = \frac{L_x}{2}, z = 0 \right ) \\
\Delta z &= \bar u_z  \left( x = \frac{L_x}{2}, z = L_z \right ) - \bar u_z \left( x = \frac{L_x}{2}, z = 0 \right ) + L_z. 
\end{aligned}
\end{equation}
Upon substituting Eqs.~(\ref{eq:solx}) and (\ref{eq:soly}) into Eq.~(\ref{eq:sector_calc}), we obtain, as a function of $\alpha_0$, the cone dimensions, and the bulk modulus $B = \mu + \lambda$, 
\begin{eqnarray}
\Delta x &=& \frac{\alpha_0}{2B} \frac{L_x L_z}{2} \\
\Delta z &=& - \frac{\alpha_0}{2B} \frac{L_z^2}{2} + L_z.
\end{eqnarray}
The sector angle is then given by, using Eq.~(\ref{eq:Phi_def}),
\begin{eqnarray}
\tan \frac{\Phi}{2} = \frac{\frac{\alpha_0}{2B} L_x }{2 (1 - \frac{\alpha_0}{2B}  \frac{L_z}{2})}. 
\end{eqnarray}
If the gravitational pressure is weak enough such that Eq.~(\ref{eq:small_reg}) holds, we can approximate $\Phi$ to first order in $\alpha_0$ as,
\begin{eqnarray} \label{eq:apex}
\Phi\approx \frac{\alpha_0}{2 B} L_x + O\left[ \left(\frac{\alpha_0}{2B}  \frac{ L_z}{2} \right) ^2 \right],
\end{eqnarray}
plotted as red dashed lines in Fig.~\ref{fig:grav}c. 
The apex angle (Fig.~\ref{fig:grav}d) in this weak pressure regime is then given by $\theta = \Phi/\pi$ as,
\begin{eqnarray} \label{eq:apex_elas}
\theta \approx \frac{\alpha_0}{2 \pi B} L_x.  
\end{eqnarray}

We now determine the smoothly varying lattice spacing $a(z)$ produced by this cone angle, using Eq.~(\ref{eq:soly}), as,
\begin{eqnarray}  \label{eq:az}
a(z) &=& a_0 + \bar u_x(x=a_0, z) - \bar u_x(x=0, z) \\
&=& a_0 \left[ 1 + \frac{\alpha_0}{2 B}(z-L_z) \right],\label{eq:ax_orig}
\end{eqnarray}
where $a_0$ is the lattice constant of the truncated cone at its upper edge, i.e. height $z = L_z$. 
The vertical coordinate $z$ of the unrolled flat sheet is proportional to the longitudinal coordinate $r$ climbing up along the surface of the truncated cone (see Fig.~\ref{fig:grav}b). When the gravitational pressure is weak enough such that the approximation in Eq.~(\ref{eq:apex}) holds, we can rewrite Eq.~(\ref{eq:ax_orig}) as,
\begin{eqnarray} \label{eq:a_sys}
a(r) = a_0  \left[ 1 + \theta \frac{\pi}{L_x} (r - r_0) \right], 
\end{eqnarray}
where $r_0$ is the total length of the cone along its inclined surface, as indicated in Fig.~\ref{fig:grav}b, and we have written $r_0 = L_z ( 1 - \theta \frac{\pi}{2} \frac{L_z}{L_x} ) \approx L_z $, an approximation accurate to first order in $\theta = \Phi / \pi$. 

The spatially varying lattice spacing in Eq.~(\ref{eq:a_sys}), derived using the elasticity solution in Eq.~(\ref{eq:soly}), is precisely that of a cylindrical lattice deformed isotropically onto the surface of a cone with apex angle $\theta$, given by,
\begin{eqnarray}
\frac{a(r)}{a_0} &=& \frac{R(r)}{R_0} = \frac{r \sin (\theta/2) }{L_x/2 \pi} \\
& \approx & 1 + \theta \frac{\pi}{L_x} ( r- r_0), \label{eq:acone}
\end{eqnarray}
where $R \equiv r \sin \frac{\theta}{2}$ is the radius of the horizontal cross section of the cone (see Fig.~\ref{fig:grav}d), and $R_0$ is the cross-sectional radius at the top of the truncated cone. 
Finally, Eqs.~(\ref{eq:soly}) and (\ref{eq:a_sys}) show that the lattice constants at the top of the cone are not distorted, since the gravitational pressure $-2 \alpha_0 (L_z - z) u_{ii}$ vanishes there in our model, as is reasonable because there are no particles above this top rim exerting gravitation forces. Hence, the circumference $L_x$ of the cylindrical rim is equal to the circumference of the top end of the truncated cone. 

Finally, we note that in arriving at the solutions above, we have assumed \textit{constant} effective coupling coefficients $\mu$ and $\lambda$. In the case of interaction potentials $V_\text{int}(r)$ which depend on the interparticle distance, inhomogeneity in the ground state lattice would render the Lam\'{e} coefficients space-dependent. This nonlinear feedback would manifest in Eq.~(\ref{eq:F_grav_cyl}) as,
\begin{eqnarray} \label{eq:F_grav_nonlinear}
F = \frac{1}{2} \int d^2 r  \left( 2 \mu (\vec r) u_{ij}^2 + \lambda(\vec r)  u_{ii}^2  + 2 \alpha_0 (L_z-z) u_{ii} \right). 
\end{eqnarray}However, we show in Appendix \ref{app:force} that this nonlinearity can be neglected in the small angle regime (i.e. Eq.~(\ref{eq:small_reg})), where the ground state equilibrium displacements are well-approximated by those provided by the linear solutions in Eqs.~(\ref{eq:solx}) and (\ref{eq:soly}). However, as the numerical calculations in Sec.~\ref{sec:phonons} show, this inhomogeneity in the elastic coefficients have a \textit{non-negligible} effect on the normal modes and is key to the physics of fluctuations. 

\subsection{Connection between elasticity theory and a microscopic model \label{sec:conn}}

We now further detail the mapping between the microscopic model in Sec.~\ref{sec:micro} and the elastic free energy in Sec.~\ref{sec:elas}. 
In the following sections, we express the pressure coefficient $\alpha_0$ and elastic constants $\mu$, $\lambda$ in Eq.~(\ref{eq:F_grav_cyl}), and the cone angle $\theta$ in Eq.~(\ref{eq:apex_elas}), in terms of the parameters of the specific microscopic Hamiltonian Eq.~(\ref{eq:H_micro}), using a Lennard-Jones pair potential. 

\subsubsection{Gravitational pressure}

Upon integrating the gravitational part of the continuum free energy Eq.~(\ref{eq:F_grav_cyl}) by parts, we obtain,
\begin{eqnarray}
F_\text{ext} &=& \alpha_0  \int d^2 r    (L_z-z) u_{ii}  \\
&=& - \alpha_0 \int d^2 r    \left[ \partial_i (L_z-z) \right]u_{i}(\vec r)  + S \label{eq:F_S}. 
\end{eqnarray}
The surface term $S$ vanishes, as can be seen by evaluating
\begin{eqnarray}
S &= & \alpha_0 \left[ \int dz (L_z-z) u_x \Big |_{x=0}^{x=L_x} + \int dx (L_z-z) u_z \Big|_{z=0}^{z=L_z} \right] \\
&= & \alpha_0 \left\{ \int (L_z-z)  \left [u_x (x=L_x, z) - u_x(x=0, z) \right] dz  \right. \\
&& \left. + L_x \left [ (L_z-L_z) u_z(x, z = L_z) -  (L_z-0) u_z(x, z = 0)\right] \right\} \notag 
\end{eqnarray}
The first term vanishes by our cylindrical boundary conditions along $x$ and the last term vanishes by the fixed boundary condition at $z =0$. The free energy due to the external potential then becomes,
\begin{eqnarray}
F_\text{ext} &=&  \alpha_0 \int d^2 r   \delta_{iz} u_{i}(\vec r) \\
% &=& \alpha_0 \int d^2 r   u_{z}(\vec r)  = \alpha_0 a^2  \sum_n  u_{n, z}\\
&=& \alpha_0 a^2  \sum_n  z_n + \text{const.} \label{eq:Fext_dis} 
\end{eqnarray}
Upon identifying,
\begin{eqnarray} \label{eq:alpha0_g}
\alpha_0 = \frac{mg}{a^2},
\end{eqnarray}
where $g$ is the gravitational acceleration and $m$ is the effective mass of one particle, we see that $F_\text{ext}$ corresponds to $H_\text{ext}$ from the microscopic Hamiltonian in Eq.~(\ref{eq:V_ext}), up to an additive contribution to the energy.

\subsubsection{Effective Lame coefficients \label{sec:lame_eff}}

To find the effective elastic coefficients corresponding to the microscopic Hamiltonian such as Eq.~(\ref{eq:H_micro}), we first recall that the dispersion relations at long wavelengths for a homogeneous lattice with short range interactions is given by elasticity theory as~\cite{ashcroft1978solid},
\begin{eqnarray}
\Omega_T = m \omega_T^2 &= & \mu (q a)^2, \label{eq:elas_T}\\
\Omega_L =m \omega_L^2 &=&  \left(2  \mu + \lambda\right) (q a)^2 \label{eq:elas_L}
\end{eqnarray}
where $q$ is the momentum, $\omega_L$ and $\omega_T$ are the longitudinal and transverse phonon frequencies, respectively, and $\Omega_L$ and $\Omega_T$ are the corresponding eigenenergies. 
As shown later in Sec.~\ref{sec:high_phonons_BE} (see also Appendix \ref{sec:power_law}),  for particles in a triangular lattice interacting via the Lennard-Jones potential in Eq.~(\ref{eq:LJ}), the dispersion curves at long wavelengths are,
\begin{eqnarray}
m \omega^2_T &=&   (qa)^2 \varepsilon \frac{27}{a^2}  \label{eq:disp_T}\\
m \omega^2_L &=&  (qa)^2 \varepsilon  \frac{3 \cdot 27}{a^2}. \label{eq:disp_L}
\end{eqnarray}Thus, the effective Lam\'{e} coefficients for our microscopic Lennard-Jones model are given by,
\begin{eqnarray}
\mu &=& \lambda = 27 \frac{\varepsilon}{a^2}  \label{eq:eff_lame}
\end{eqnarray}
We immediately see that the elastic constants are spatially varying if the lattice spacing $a=a(r)$ is spatially varying, as in the case of the cone. 
Note that the corresponding Poisson ratio $\nu = \lambda/(2 \mu + \lambda)$ is equal to $1/3$, which satisfies the Cauchy condition for isotropic interaction potentials~\cite{weiner2012statistical}. 
(We note that for systems of particles with sufficiently long-range repulsive interactions such as $V_\mathrm{int} (r) \sim 1/r$, the effective $\lambda$ becomes infinite in the long wavelength limit~\cite{fisher1979defects,fisher1980flux}. This is equivalent to the observation that the Wigner crystal is incompressible~\cite{fisher1979defects}.) 

\subsubsection{Cone Angle}

Upon combining Eqs.~\ref{eq:apex}, \ref{eq:alpha0_g}, and \ref{eq:eff_lame}, we obtain the apex angle of the cone in terms of the microscopic Lennard-Jones pair potential parameter $\varepsilon$ as,
\begin{eqnarray} \label{eq:apex_micro}
\theta = \frac{L_x mg}{108 \pi \varepsilon}.
\end{eqnarray}
Thus, under a gravitational field parallel to the axis of symmetry, a cylindrical shell of circumference $L_x$, consisting of particles with effective mass $m$ interacting via the LJ potential in Eq.~(\ref{eq:LJ}) with strength $\varepsilon$, will assume the shape of a cone with angle $\theta$ given by Eq.~(\ref{eq:apex_micro}). 

\section{Cone Phonons: Frequency-dependent Localization \label{sec:phonons}}

Can you hear the shape of a cone? In this section, we examine the eigenmodes of a 2d lattice in the shape of a truncated cone. We find that the cone shape, namely its apex angle $\theta$, controls level crossings at long wavelengths and eigenfunction quasi-localization at short wavelengths. 

In Sec.~\ref{sec:phonon_calc}, we study the fluctuation energy to second order in particle displacements. 
In Sec.~\ref{sec:low_phonons}, we show that the low frequency normal modes have robust vortex and anti-vortex configurations in their eigenfunctions, and exhibit changes in energy ordering as a function of the cone angle $\theta$.  

\begin{figure*}
    \centering
    \includegraphics[width=0.95\textwidth]{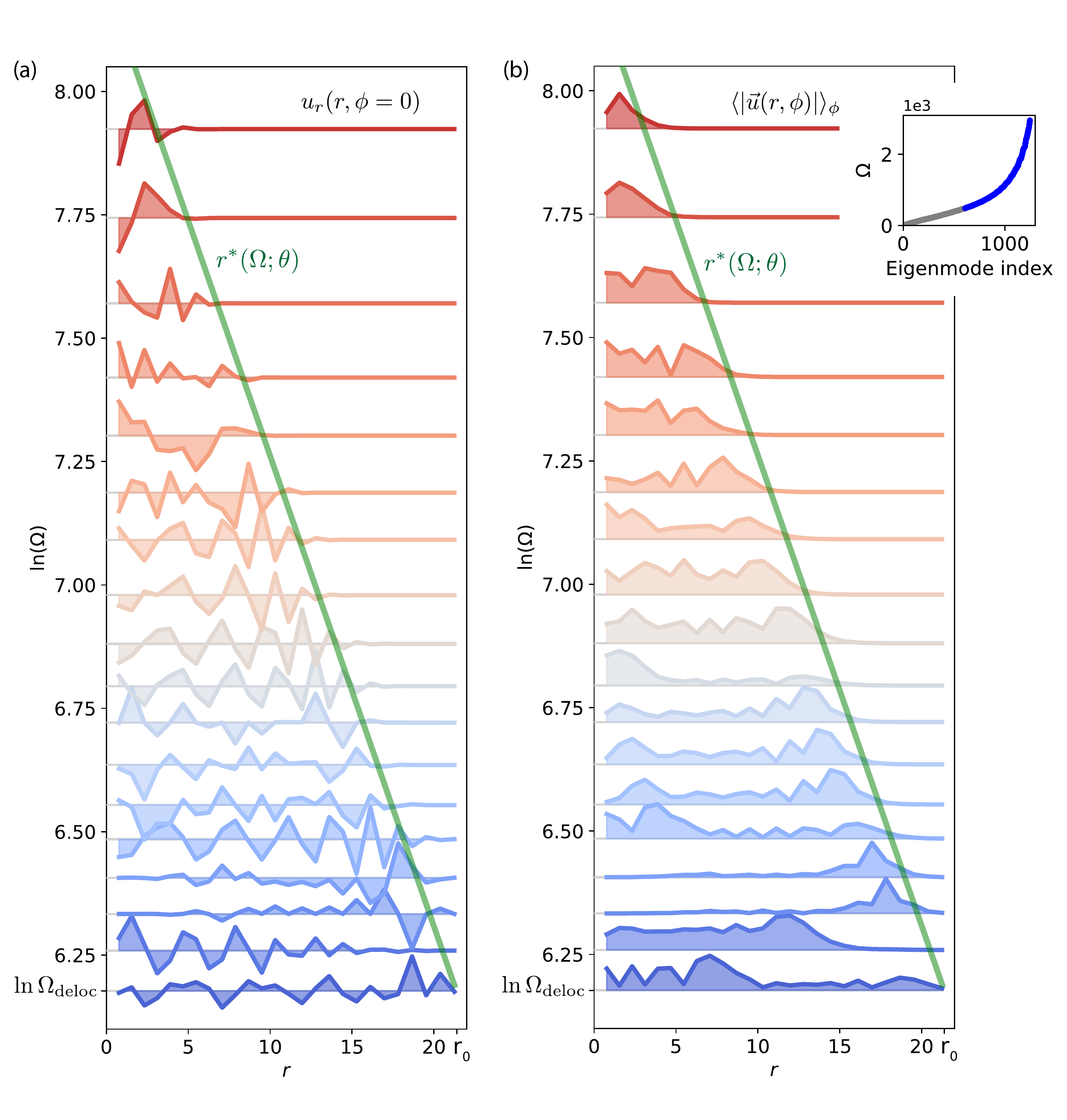}
    \caption{Localization of high frequency eigenfunctions for $N$ particles interacting via a LJ potential on the surface of a cone, with geometrical parameters $(L_z, L_x,  N,\theta) = (13 \sqrt{3}, 24, 648, 1.1^\circ)$, longitudinal cone length $r_0 \approx 21.25$, and undeformed lattice constant $a_0 = 1$. (a)-(b): The longitudinal component (a) and absolute magnitude (b) of the numerical eigenfunctions stacked up vertically according to the logarithm of their eigenenergies. In (a), $u_r(r, \phi = 0)$ is the longitudinal displacement at one value of $\phi$ and $\langle |\vec u (r, \phi) | \rangle_\phi $ is the displacement magnitudes in each row of the cone averaged over the azimuthal angle $\theta$ (e.g. for lattice sites with the same $r$ coordinate). The green line is Eq.~(\ref{eq:loc_fin}), which clearly identifies the localization domain boundaries of the eigenmodes as a function of $\Omega$. $\Omega_\mathrm{deloc}$ marks the eigenenergy below which all eigenfunctions are delocalized. The inset on the right plots the eigenenergies $\Omega$ as a function of the eigenmode index. The blue points highlight the range of eigenenergies whose corresponding eigenmodes experience quasi-localization.}
    \label{fig:LJ_cone}
\end{figure*}

In the upper half of the phonon spectrum, we find that eigenmodes at high frequencies are confined towards the bottom of the cone, where the lattice is most compressed.
We use two methods to identify the analytical relation between the eigenenergy $\Omega$ and the position $r^*$ of the localization domain boundary, measured from the base up the flanks of the cone. In Sec.~\ref{sec:high_phonons_BE}, we find the space-dependent local band edge (the maximum eigenenergy above which all modes decay exponentially) by calculating the dispersion relation. In Sec.~\ref{sec:high_phonons_WKB}, we map our equations of motion onto Schrodinger's equation and find the turning point using an upside-down WKB analysis. These two methods give identical results: the localization domains $[0, r^*]$ of the high-frequency eigenfunctions on a cone with small apex angle $\theta$ are given, in terms of the parameters $\varepsilon$ and $a_0$ specifying the Lennard-Jones pair potential (Eq.~(\ref{eq:LJ})), by, 
\begin{eqnarray} \label{eq:loc_fin}
r^*(\Omega; \theta) &= r_0 + \frac{3}{31 \pi} \frac{L_x}{\theta}\left[ C(\varepsilon, a_0) - \ln \Omega \right] 
\end{eqnarray}
where $r_0$ is the total length along the cone flank, $C(\varepsilon, a_0) = 432 \varepsilon / a_0^2$ is a constant, and $L_x$ and $a_0$ are the circumference and lattice constant, respectively, of the undeformed cylinder. (The parameter $a_0$ is also the lattice constant along the top rim of the truncated cone where the lattice is undeformed by the gravitational field.) Thus, given the phonon eigenenergy $\Omega$, Eq.~(\ref{eq:loc_fin}) predicts the location $r^*$above which its eigenfunction decay exponentially. Note that $r$ and $r^*$ in Eq.~(\ref{eq:loc_fin}) can be replaced with $z$ and $z^*$ to first order in $\theta$. Finally, Eq.~(\ref{eq:loc_fin}), evaluated when $r^* = r_0$ also gives the threshold eigenfrequency $\Omega_\text{deloc}$ below which all eigenfunctions are delocalized,
\begin{eqnarray}
\ln \Omega_\text{deloc} = C(\varepsilon, a_0). 
\end{eqnarray}
The criterion embodied in Eq.~(\ref{eq:loc_fin}) is plotted in green in Fig.~\ref{fig:LJ_cone} and shows good agreement with the numerical eigenfunctions.

\subsection{Phonon spectrum \label{sec:phonon_calc}}

To examine the fluctuations about the conical ground state in Figs.~\ref{fig:grav}b and c, we decompose the position of the $n$-th particle as $\vec r_n = \vec R_n + \vec u_n$, where $\vec R_n$ is its equilibrium position and $\vec u_n$ is its displacement away from equilibrium. 
Hence forth, $i,j, k = 1,2$ will be used as direction indices and $n, m = 1, \cdots N$ as particle indices. 

% Upon redefining the displacement $u_i \rightarrow u_i - \bar u_i$ about the equilibrium configuration in Fig.~\ref{fig:grav}c (Eqs.~\ref{eq:solx} and \ref{eq:soly} with cylindrical boundary conditions)  and 
Upon expanding the fluctuation energy to second order in displacements $\vec u_n$, the linear terms vanish by force balance. The gravitational potential, although it determines the spatial variation of the lattice constant in the ground state, drops out. 
On neglecting constant terms, we obtain the fluctuation energy as,
\begin{eqnarray} \label{eq:H_dis_full}
\Delta E = \sum_{n \neq m} \frac{1}{2} \Pi_{ij}(\vec R_n - \vec R_m) (u_n - u_m)_i  (u_n - u_m)_j
\end{eqnarray}
where, for the Lennard-Jones interaction potential in Eq.~(\ref{eq:LJ}), $\Pi_{ij}(\Delta \vec R)$ is given by,
\begin{eqnarray} \label{eq:Pi_ij}
\Pi_{ij}(\Delta \vec R) &=& \varepsilon \left[ a_0^{12} \left( 12 ( 12 + 2) \frac{\Delta R_i \Delta R_j}{|\Delta \vec R|^{12+4}} - 12\frac{\delta_{ij}}{|\Delta \vec R|^{12+2}} \right)  \right . \notag \\
& & \left. - 2 a_0^{6} \left( 6 ( 6 + 2) \frac{\Delta R_i \Delta R_j}{|\Delta \vec R|^{6+4}} - 6\frac{\delta_{ij}}{|\Delta \vec R|^{6+2}} \right) \right]  
\end{eqnarray}

As a check, we will compare our analytical results in Sec.~\ref{sec:high_phonons_BE} and \ref{sec:high_phonons_WKB} against numerical eigenfunctions. These eigenfunctions are obtained by solving the real-space eigenproblem,
\begin{eqnarray} \label{eq:eig_prob}
\mathbf{M} {\bf u}^{(\alpha)} = \Omega_\alpha {\bf u}^{(\alpha)},
\end{eqnarray}
with fixed boundary conditions at the lower rim, where $\alpha$ indicates the $\alpha$-th normal mode and the dynamical matrix $\mathbf{M}$ depends on $\Pi_{ij}$ (see Appendix \ref{app:M_calc} for details). 
% We note that the localization phenomena we observe in the high frequency phonons is not sensitive to the top and bottom boundary conditions 
% Additionally, we observe that the eigenmodes for the top clamped cone and the bottom clamped cone do not show significant differences in terms of localization. 

\subsection{Properties of low frequency eigenmodes \label{sec:low_phonons}}

\begin{figure*}[htb]
    \centering
    \includegraphics[width=1\textwidth]{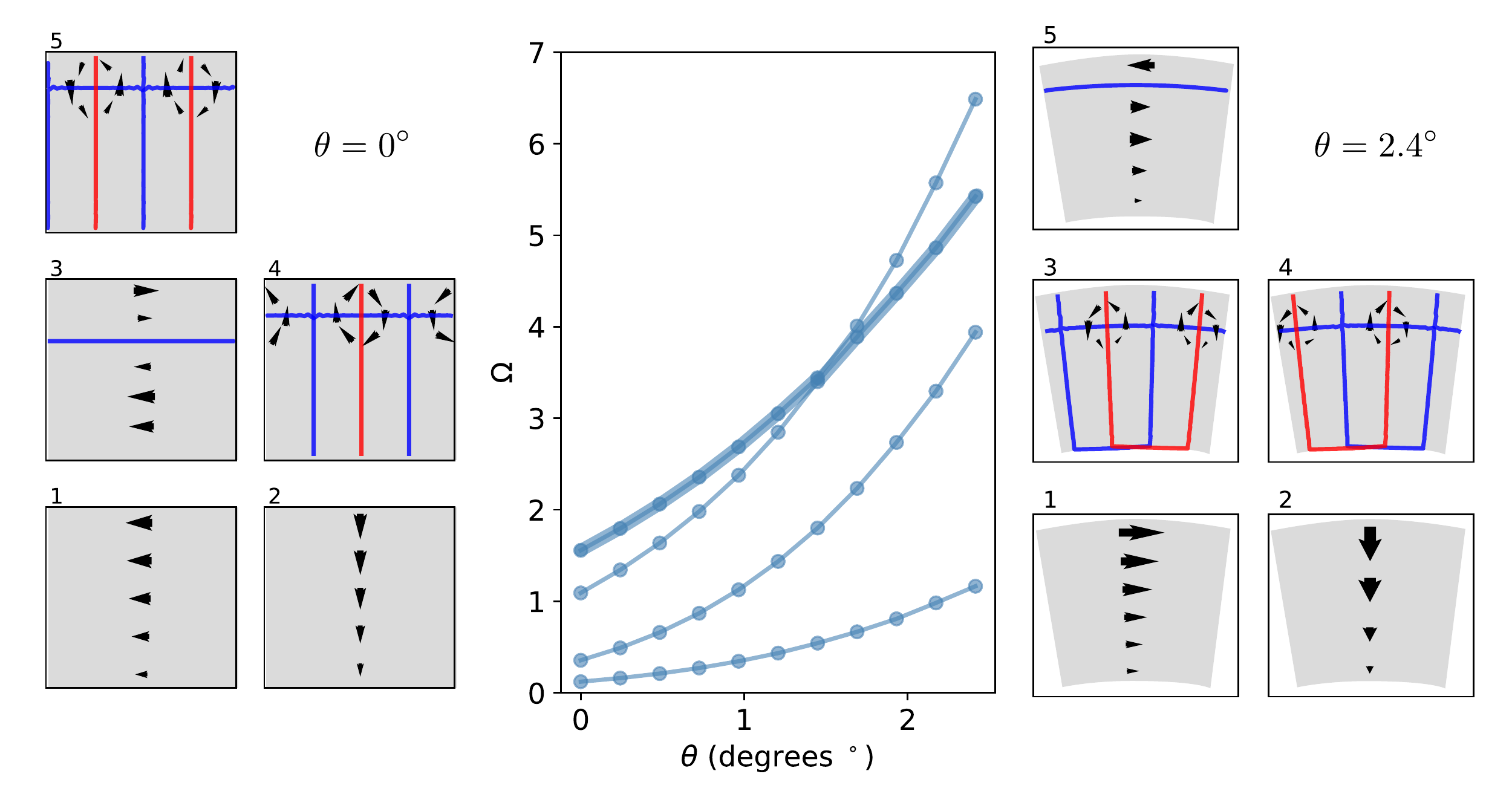}
    \caption{The central plot shows the five lowest energy levels of the cone as a function of increasing apex angle $\theta$. The corresponding eigenmodes are shown in the cylindrical limit $\theta = 0^\circ$ (left) and at $\theta = 2.4^\circ$ (right). The enlarged black arrows indicate the displacements at selected lattice sites. The blue and red lines indicate zeros of the horizontal displacements $u_x$ and vertical displacements $u_z$, respectively. The level crossing between the  (T, $n_r = 1, n_\phi = 0$) mode and the (T, $n_r = 1, n_\phi = 1$),  (T, $n_r = 1, n_\phi = -1$) doublet (shown as the thick line in the central plot) occurs at $\theta \approx 1.5^\circ$. As discussed in the text $T$ specifies a \textit{transverse} long wavelength mode, while $n_r$ and $n_\phi$ are integers specifying node numbers.}
    \label{fig:cross}
\end{figure*}

The low frequency normal modes exhibit interesting behavior as a function of the apex angle $\theta$. Fig.~\ref{fig:cross} shows the the five lowest eigenenergies and their eigenfunctions at $\theta = 0$ (the limit of the cylinder)  and $\theta \approx 2.42^\circ$. We can classify the eigenfunctions by node-counting. The blue lines mark where the horizontal displacement $u_x$ crosses 0 and the red lines mark where the vertical displacement $u_z$ crosses 0. 
Note that the crossing of two longitudinal node lines of different orientations (i.e. a horizontal blue ($u_x = 0$) line and a vertical red ($u_z = 0$) line) coincides with a vortex-like texture in the displacement field, while the crossing of two transverse node lines of difference orientations (i.e. a vertical blue ($u_x = 0$) line and a horizontal red ($u_z = 0$) line) corresponds to an anti-vortex configuration in the displacement field. These vortices and anti-vortices in the eigenfunctions are stable to perturbations, or in other words, topologically protected. These vortex configurations are thus preserved as we tune the apex angle away from 0. In contrast, the crossing between a transverse node line and a longitudinal node line (i.e. two red lines or two blue lines) are not robust and can easily vanish under a slight conical deformation. 

Since the vortex configurations stay robust under the conical perturbation, we can use them to identify and track the eigenmodes as we tune $\theta$ adiabatically. For a cylinder $(\theta = 0)$, the number of nodes along the $x$ and $z$ directions, $n_x$ and $n_z$, determine the wave numbers of the eigenfunctions at long wavelengths. For periodic boundary conditions along $x$ and bottom clamped boundary condition along $z$, these wavefunctions behave according to,
\begin{eqnarray}
u (x,z) \sim \sin (k_z z) \cos (k_x x)
\end{eqnarray}
where,
\begin{equation}
\begin{aligned}
k_z =& \frac{\pi}{L_z} (n_z + \frac{1}{2}), \quad &n_z = 0, 1, 2, \cdots \\
k_x =& \frac{2\pi}{L_x} n_x, \quad &n_x = 0, \pm 1, \pm2, \cdots
\end{aligned}
\end{equation}
Note that in cone coordinates, we have $n_z \rightarrow n_r$ and $n_z \rightarrow n_\phi$. 
Additionally, since we are in two dimensions, the eigenmodes at low energies split into two branches, transverse and longitudinal (see Fig.~\ref{fig:LJ_disp}). To combine this information, we can classify each low energy normal mode by identifying whether it is transverse (T) or longitudinal (L), and then counting the number of nodes along $r$ ($n_r$) and $\phi$ $( n_\phi)$. 
For example, in the cylindrical limit $\theta = 0$, the eigenmodes, in order of increasing frequency, are: (T, $n_r = 0, n_\phi = 0$),  (L, $n_r = 0, n_\phi = 0$),  (T, $n_r = 1, n_\phi = 0$),  (T, $n_r = 1, n_\phi = 1$),  (T, $n_r = 1, n_\phi = -1$). 

In contrast, at $\theta =2.42^\circ$, this ordering changes to: (T, $n_r = 0, n_\phi = 0$),  (L, $n_r = 0, n_\phi = 0$),  (T, $n_r = 1, n_\phi = 1$),  (T, $n_r = 1, n_\phi = -1$), (T, $n_r = 1, n_\phi = 0$). That is, eigenmode 3 swaps order with eigenmodes 4 and 5. As shown in Fig.~\ref{fig:cross}, this level crossing occurs at approximately $\theta \approx 1.52^\circ$.
Note that eigenmodes 4 and 5 at $\theta= 0 $ (eigenmodes 3 and 4 after the level crossing) are degenerate doublets. 

This level crossing is interesting for two reasons. First, it suggests that we can probe the angle of the cone by measuring the ordering of its lowest energy eigenmodes.
Second, we know that for Hermitian matrices, level crossings for a system with $k$ parameters can only occur on a $(k-2)$ manifold~\cite{mehta2004random,wigner1967random,landau2013quantum}. Since our conical system has only one parameter ($\theta$), level crossings are not allowed unless some symmetry is present to relax the condition. What accidental symmetry in this system allows the eigenvalues to cross when $\theta \approx 1.5^\circ$, is an interesting question for future investigations.

\subsection{Moving band edges \label{sec:high_phonons_BE}}

In this section, we derive the localization domain boundary of the \textit{high} frequency phonons on the cone by first calculating the band edge of the dispersion relations for the interacting particles on a uniform triangular lattice. Provided the lattice parameter varies slowly in space (i.e. if the external gravitational field is weak), we can compute a \textit{local} band edge that varies spatially throughout the cone as a function of $r$, giving us a stopping criterion for a fixed phonon eigenenergy $\Omega$. 
We compare the predicted domain boundary to numerical eigenfunctions and see good agreement. 

Since the Lennard-Jones (LJ) interaction potential is short ranged, we truncate to nearest neighbor interactions in the following calculations. (As detailed in Appendix~\ref{sec:power_law}, the addition of next nearest neighbor interactions has a negligible effect on the band edge.) Upon rewriting the summation in Eq.~(\ref{eq:H_dis_full}) as a sum over center sites $\{\vec R_m \}$ and their nearest neighbors $\{ \vec R_m + \vec n_\alpha | \alpha = 1, \cdots, 6\}$, where $\{\vec n_\alpha \}$ are the 6 nearest neighbor lattice vectors, we obtain,
% \begin{eqnarray}
% \Delta E &=& \frac{1}{2} \sum_m \sum_{ij} \sum_{\alpha = 1, \cdots, 6} \Pi_{ij} (\vec n_\alpha) \left( u_i (\vec R_m + \vec n_\alpha)   u_j (\vec R_m + \vec n_\alpha)\right. \notag \\
% &&\quad \left.  +u_i (\vec R_m) u_j (\vec R_m)   - u_i u_i (\vec R_m + \vec n_\alpha) u_j (\vec R_m) - u_j u_j (\vec R_m) u_i (\vec R_m) \right). 
% \end{eqnarray}
\begin{equation} \label{eq:E_nn}
\begin{aligned}
\Delta E &= \frac{1}{2} \sum_m \sum_{ij} \sum_{\alpha = 1, \cdots, 6} \Pi_{ij} (\vec n_\alpha) \left( u_i (\vec R_m + \vec n_\alpha) -u_i (\vec R_m)  \right) \\
& \times \left(  u_j (\vec R_m + \vec n_\alpha) - u_j (\vec R_m)  \right)  
\end{aligned}
\end{equation}
After incorporating the dynamical term $\sum_n \frac{m}{2} |\dot {\vec u}_n|^2 $ in the Hamiltonian, calculating the equations of motion, and assuming a wave-like solution of the form,
\begin{eqnarray}
u_i(\vec R_m) = e^{i \vec q \cdot \vec R_m} e^{i \omega t} u_i(\vec q),
\end{eqnarray}
where $\vec q$ is the momentum and $\omega$ is the physical frequency of the in-plane phonons, we arrive at the following eigenvalue problem,
\begin{eqnarray}
\Omega(\vec q) u_i( \vec q) = D_{ij}(\vec q) u_j (\vec q)
\end{eqnarray}
% \begin{eqnarray}
% u_{m, i} = \int \frac{d^2 q}{(2 \pi)^2} e^{i \vec q \cdot \vec R_m} u_i(\vec q),
% \end{eqnarray}
% the fluctuation energy becomes
% \begin{eqnarray}
% \Delta E = \int \frac{d^2 q}{(2 \pi)^2} D_{ij} (\vec q) u_i(\vec q) u_j (-\vec q),
% \end{eqnarray}
where the dynamical matrix $D_{ij} (\vec q)$ is given by,
\begin{eqnarray}
D_{ij} (\vec q) = \sum_{\alpha = 1, \cdots, 6} \Pi_{ij} (\vec n_\alpha) \left (1 - \cos(\vec n_\alpha \cdot \vec q)\right)\label{eq:Dij}
\end{eqnarray}
and $\Omega (\vec q) = m \omega(\vec q)^2$ are its eigenvalues. 

Calculation of the dynamical matrix and its eigenvalues is detailed in Appendix \ref{sec:power_law}. Fig.~\ref{fig:LJ_disp} shows the analytical dispersion curves in the first Brillouin zone. Note that, interestingly, upon traversing a single band in a continuous loop in momentum space $\Gamma-K-M-\Gamma$ , starting from and circling back the zone origin $\Gamma$, a transverse branch (orange) transforms smoothly into a longitudinal branch (blue) upon returning to $\Gamma$, and vice versa. 
% As stated in Sec.~\ref{sec:lame_eff}, we can now calculate the dispersions near the Brillouin zone center (Eqs.~\ref{eq:disp_T} and \ref{eq:disp_L}) and compare them to Eqs.~\ref{eq:elas_T} and \ref{eq:elas_L} to extract the effective Lame coefficients in Eq.~\ref{eq:eff_lame}. 

\begin{figure}[htb]
    \centering
    \includegraphics[width = 0.95\columnwidth]{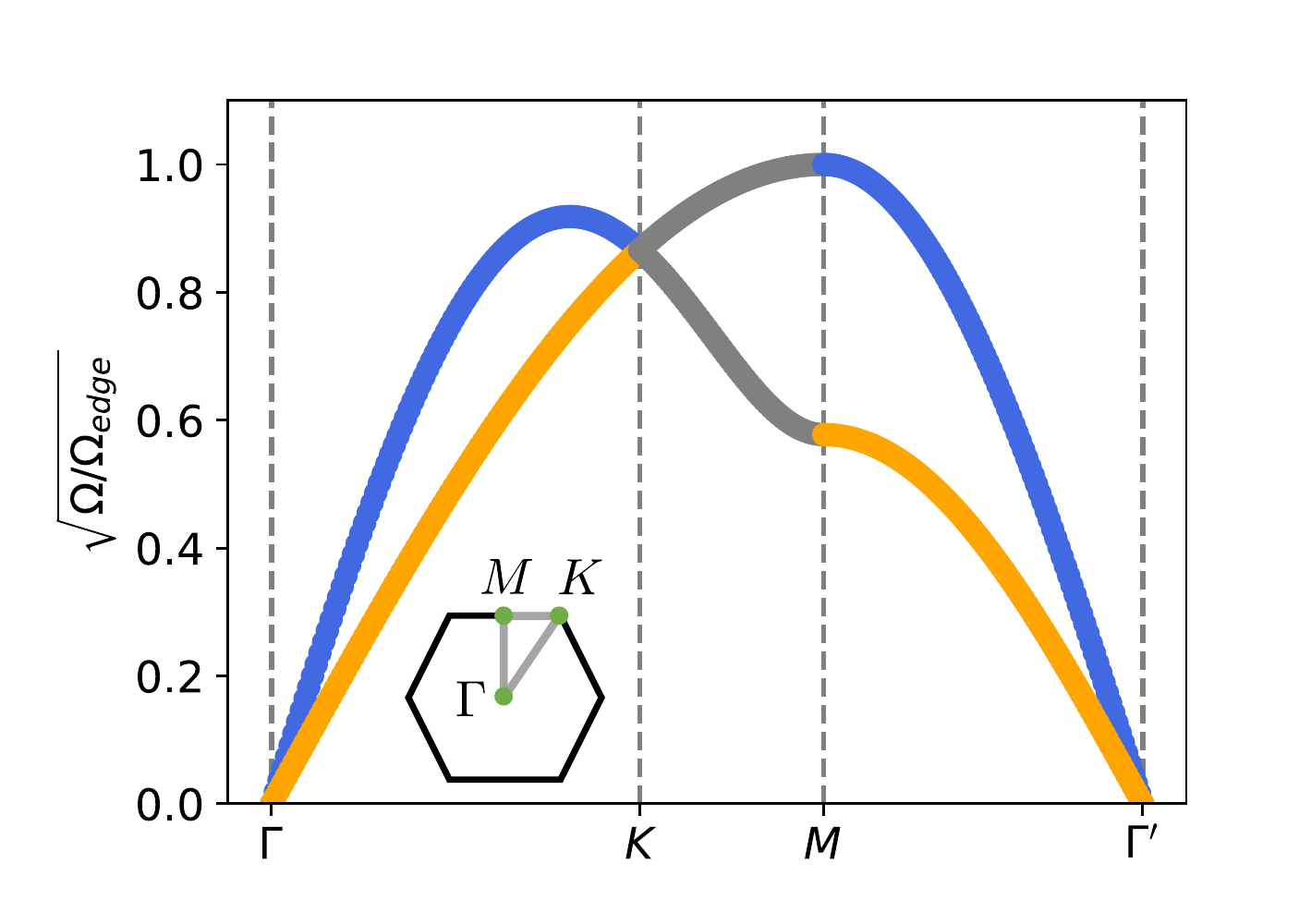}
    \caption{Phonon dispersion curves for a uniform triangular lattice resulting from nearest neighbor interactions under the Lennard-Jones potential along the $\Gamma-K-M-\Gamma'$ loop of the first Brillouin zone, as shown in the inset. Near the Brillouin zone center $\Gamma$, lower phonon branches have transverse characteristics (orange) and higher phonon branches have longitudinal characteristics (blue). High frequency modes along the Brillouin zone boundary have mixed transverse and longitudinal properties (gray). Note that as one moves from left to right along the blue-gray-orange path on the circuit above, a longitudinal phonon excitation smoothly changes into a transverse mode. A similar transformation takes place along the orange-gray-blue path from $\Gamma$ to $\Gamma'$. }
    \label{fig:LJ_disp}
\end{figure}

As shown in Appendix \ref{sec:power_law}, the upper band edge of the spectrum (i.e. the symmetry point $M$ in Fig.~\ref{fig:LJ_disp}) for a uniform lattice with an LJ pair potential and lattice constant $a_0$ is simply,
\begin{eqnarray} \label{eq:BE_LJ}
\Omega_\text{edge}^{} &=&  432 \frac{\varepsilon}{a_0^{2}}.
\end{eqnarray}
For a slowly varying inhomogeneous lattice with $a = a(r)$, we expect that the band edge becomes space-dependent as,
\begin{eqnarray} \label{eq:BE_LJ_NN}
\Omega_\text{edge}^{}[a(r)] &=& \frac{6 \varepsilon}{a(r)^2}\left[ 152 \left( \frac{a_0}{a(r)} \right)^{12}  -  80 \left( \frac{a_0}{a(r)} \right)^6\right].
\end{eqnarray}
Upon substituting Eq.~(\ref{eq:a_sys}) for the lattice spacing $a(r)$ of the cone into the band edge criterion in Eq.~(\ref{eq:BE_LJ_NN}) and expanding to first order in $\theta$, we recover the localization condition for small cone angles displayed in Eq.~(\ref{eq:loc_fin}).

\subsection{Effective potential and Inverted WKB \label{sec:high_phonons_WKB}}

In this section, we obtain further insights by identifying a change of variable that smooths out the rapidly varying displacement field of the phonon at the band edge. This transformation allows us to map our eigenvalue equation in the continuum limit onto a Schrodinger-like equation with the momentum transformation $p \rightarrow ip$. We then apply an inverted 1d WKB analysis to the cone along the $r$ direction to arrive at an alternative derivation of Eq.~(\ref{eq:loc_fin}).

\begin{figure}[htb]
    \centering
    \includegraphics[width = 1\columnwidth]{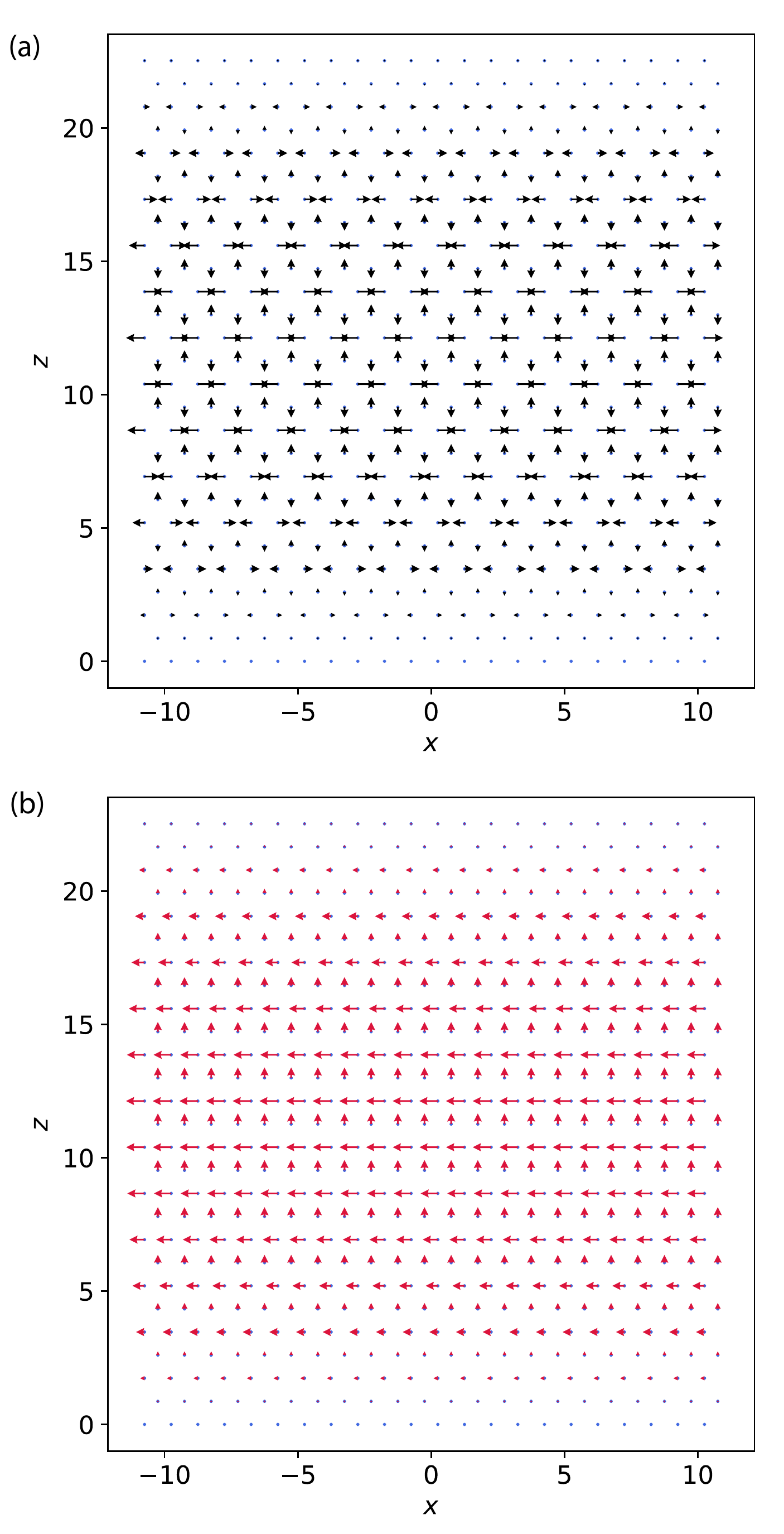}
    \caption{The displacements corresponding to the highest energy eigenmode of the cylinder with a free boundary at the top and a fixed boundary at the bottom before (a) and after (b) the change of variables illustrated in Fig.~\ref{fig:trans}a.}
    \label{fig:cyl_bandedge}
\end{figure}

\begin{figure}[h!]
    \centering
    \includegraphics[width = 0.95\columnwidth]{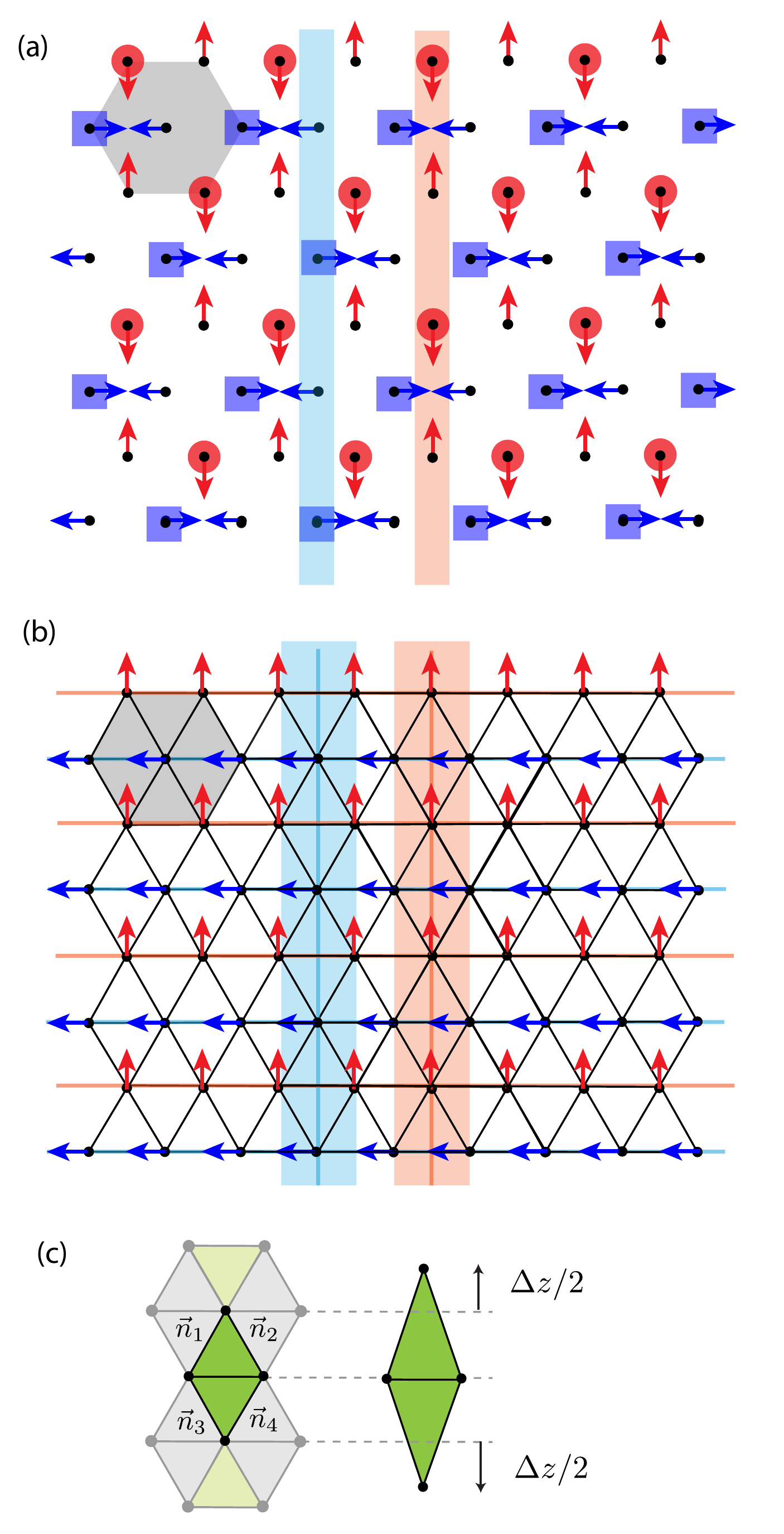}
    \caption{(a): Displacement configuration of the highest frequency eigenmode for the cylindrical shell. The presence of a colored shape at a site indicates that the sign of the displacement there is flipped according to our change of variables. Red circle (blue square) indicates that the flipped displacement at that site is purely vertical (horizontal). Note that along each vertical stripe, the displacement at every other site is flipped.
(b): The red (blue) vertical stripe highlights the sites aligned in the horizontal direction, interleaved by rows where there are zero vertical (horizontal) displacements. We coarse grain over each interleaving row by extracting an effective spring constant. 
 (c): Vertical stretching of a pair of sites aligned in the horizontal direction by $\Delta z$. The effective spring constant $k_\text{1d}$ is extracted from the elongation energy of the four affected bonds $\{\vec {n}_b | b = 1, \cdots, 4 \}$. }
\label{fig:trans}
\end{figure}

Figure \ref{fig:cyl_bandedge}a shows the highest frequency eigenmode for the cylinder. The displacement field varies rapidly in space on the scale of the lattice constant, and gradually vanishes near the top and bottom edges. In Fig.~\ref{fig:trans}a, we subtract the gradual modulation in the $z$ direction to resolve the short-wavelength pattern of the displacements. By applying a change of variables that flips the sign of the displacement in every other row and every other column (marked by the colored sites in Fig.~\ref{fig:trans}a), we arrive at a configuration where the displacements in every other row are approximately identical in orientation and magnitude, while the rows alternate between pure vertical displacement and pure horizontal displacements (Fig.~\ref{fig:trans}b). 

Upon applying this change of variables to the band-edge eigenfunction in Fig.~\ref{fig:cyl_bandedge}a, the transformed displacement field in Fig.~\ref{fig:cyl_bandedge}b has two key characteristics: (1) displacements within the same row are identical, and (2) displacements in every other row vary smoothly along the vertical direction. In Fig.~\ref{fig:trans}b, the red (blue) stripes highlights the sites aligned in the horizontal direction, interleaved by rows where there is zero vertical (horizontal) displacement. 

Hence, we reduce the two-dimensional eigenfunction problem to a one-dimensional one along the $z$ coordinate of the cylinder (i.e. the $r$ coordinate along the flank of the cone), as indicated by the vertical stripes in Fig.~\ref{fig:trans}b. We then ``integrate out'' the intercepting row, where the displacement direction abruptly rotates by $90^\circ$, by calculating an effective spring constant between two horizontally aligned sites. 
As shown in Fig.~\ref{fig:trans}c, to calculate this effective spring constant, we imagine stretching the vertical distance between the two sites by $\Delta z$. We then calculate the elastic energy $\Delta E_s$ from this stretching $\Delta u_i = \pm \delta_{i2} \Delta z / 2$ using Eq.~(\ref{eq:Pi_ij}) as follows,
\begin{eqnarray}
\Delta E_s(\Delta z) &=& \sum_{b = 1, \cdots, 4} \Pi_{ij}(\vec n_b) \left( \delta_{2i} \frac{\Delta z}{2} \right) \left( \delta_{2j} \frac{\Delta z}{2} \right),
\end{eqnarray}
where $\{\vec n_b \}$ consists of the four bonds (labeled in Fig.~\ref{fig:trans}c) whose lengths are elongated by the vertical stretching. Upon rewriting the elastic energy in the form of Hooke's law,
\begin{eqnarray}
\Delta E_s( \Delta z) \equiv \frac{1}{2} k_\text{1d}\left(\Delta z\right)^2, \label{eq:E_dx}
\end{eqnarray}
we extract the effective spring constant as,
\begin{eqnarray} \label{eq:kLJ}
k^{}_\text{1d} (r)&=&  \frac{3}{2} \frac{\varepsilon}{a(r)^{2}} \left[ 152\left( \frac{a_0}{a(r)} \right)^{12}  - 80 \left( \frac{a_0}{a(r)} \right)^{6}  \right].
\end{eqnarray}
where the spatially varying lattice constant $a = a(r)$ for the cone is given by Eq.~(\ref{eq:a_sys}). 

We can now model the vertical stripe in Fig.~\ref{fig:trans}b as a 1d chain of springs with the effective spring constant given in Eq.~(\ref{eq:kLJ}). The resulting Lagrangian is given by,
\begin{eqnarray} \label{eq:L_1d}
\mathcal{L}[\mathbf{u}]=\sum_{n} \left[ \frac{m}{2} \dot{u}_{n}^{2}-\frac{1}{2} k^{}_{1d}[a(r_n)]\left( u_{n}-u_{n+1}\right)^{2} \right],
\end{eqnarray}
where $u_n$ is the displacement scalar of the $n$-th site along the vertical strip. We can choose to model the red or blue strip in Fig.~\ref{fig:trans}, for which $u_n$ would represent the vertical or horizontal component of the displacement; these quantities have the same equations of motion. 
% The equations of motion are given by 
% \begin{eqnarray} \label{eq:eom_init}
% -\Omega u_n + k_n(u_n - u_{n+1}) + k_{n-1} (u_n- u_{n-1}) &=& 0. 
% \end{eqnarray} 

As highlighted in Fig.~\ref{fig:trans}a, the change of variables flips the displacement at every other site in this 1d chain: $u_{n-1}, u_{n+1} \rightarrow -u_{n-1}, -u_{n+1}$. Upon applying this change of variables, the equations of motion simplify to,
\begin{eqnarray} \label{eq:eom_init}
(-\Omega + 4 k_n) u_n + k_n(u_{n+1} - 2 u_n + u_{n-1}) &=& 0,
\end{eqnarray} 
where  $\Omega \equiv m \omega^2 $ as before, $k_n \equiv k_{1d} (r_n)$, and we have approximated $k_n \approx (k_n + k_{n-1})/2$. Upon taking the continuum limit by setting$ \partial_r^2 u(r_n) = (u_{n-1} -2 u_n + u_{n+1})/a_n^2$, where $a_n = a(r_n)$,
we arrive at the following equation of motion in the continuum limit,
\begin{eqnarray} \label{eq:WKB}
\frac{1}{2 m_{\mathrm{eff}}(r)} \partial_{r}^{2} u(r)+V_{\mathrm{eff}}(r) u(r)=\Omega u(r), 
\end{eqnarray}
where,
\begin{eqnarray}
V_{\mathrm{eff}}(r)=4 k_{1d}^{}(r), \quad m_{\mathrm{eff}}(r)=\left [2 a(r)^2 k_{1d}^{}(r)\right]^{-1}.
\end{eqnarray}

Eq.~(\ref{eq:WKB}) compares directly with the 1d time-independent Schrodinger's equation,
\begin{eqnarray} \label{eq:schr}
-\frac{\hbar^{2}}{2 m} \frac{d^{2}}{d r^{2}} \Psi(r)+V(r) \Psi(r)=E \Psi(r),
\end{eqnarray}
whose local wave vector / decay rate of the wavefunction $\Psi$ is given by WKB analysis as~\cite{wentzel1926verallgemeinerung,kramers1926wellenmechanik,brillouin1926mecanique,jeffreys1925certain,green1838motion},
\begin{eqnarray}
p=\left\{\begin{array}{ll}
\pm i \sqrt{2 m(V(r)-E)}), & \text { if } E<V(r) \\
\pm  \sqrt{2 m(E-V(r))}, & \text { if } E>V(r).
\end{array}\right.
\end{eqnarray}
However, Eq.~(\ref{eq:WKB}), compared with Eq.~(\ref{eq:schr}), has a flipped sign in front of the double derivative term, corresponding to momentum transformation $p \rightarrow ip$. 
The local wave vector / decay rate of a phonon of the cone is thus given by,
\begin{eqnarray}
p=\left\{\begin{array}{ll}
\left.\pm \sqrt{2 m_{\mathrm{eff}}(r)\left(V_{\mathrm{eff}}(r)-\Omega\right.}\right), & \text { if } \Omega<V_{\mathrm{eff}}(r) \\
\pm i \sqrt{2 m_{\mathrm{eff}}(r)\left(\Omega-V_{\mathrm{eff}}(r)\right)}, & \text { if } \Omega>V_{\mathrm{eff}}(r).
\end{array}\right.
\end{eqnarray}
Since $m_\text{eff}(r) > 0$ is always positive, the turning point $r^*$ for an eigenmode with eigenfrequency $\Omega$ is given by,
\begin{eqnarray} \label{eq:En_bound}
\Omega=V_\text{eff}\left(r^*\right) = 4 k_{1d}^{}(r^*).
\end{eqnarray}
The WKB domain boundary for the phonons of a cone is thus,
\begin{eqnarray}\label{eq:WKB_c}
\Omega = 6 \frac{\varepsilon}{a(r^*)^{2}} \left[ 152\left( \frac{a_0}{a(r^*)} \right)^{12}  - 80 \left( \frac{a_0}{a(r^*)} \right)^{6}  \right].
% \frac{6 A}{a(r_t)^{t+2}} (t^2 + \frac{2}{3} t).
\end{eqnarray}
Eq.~(\ref{eq:WKB_c}) is exactly equivalent to the band edge criterion in Eq.~(\ref{eq:BE_LJ_NN}), which again becomes Eq.~(\ref{eq:loc_fin}) in the limit of small cone angle. 
\section{Conclusion}

We demonstrate the intriguing effects smooth variations in lattice parameters can have on a discrete ordered system. In particular, we show that a two-dimensional sheet of interacting particles with cylindrical boundary conditions collapses into a truncated cone under gravitational pressure. 
We find that the cone shape, namely its apex angle $\theta$, controls level crossings at long wavelengths as well as the quasi-localization of normal modes at short wavelengths, with localization domains modulated by the inhomogeneous density profile. In the regime of small cone angles $\theta$, we predict the boundaries of these domains as a function of $\theta$ in Eq.~\ref{eq:loc_fin}.

The system studied in this paper can be realized in experiments, for example, by a two-dimensional colloidal crystal (possibly in a liquid solvent) with colloids adsorbed onto the surface of a cone~\cite{sun2020colloidal,plummer2021sticky} at the appropriate cone angle, given by Eq.~\ref{eq:apex} as a function of colloid's effective mass in the solvent. In this work, we have examined inertial dynamics $\frac{m}{2} \dot u^2$ for concreteness, but other types of (possibly overdamped) dynamics can also be studied by similar methods.

Similar localization phenomena may be found in other types of inhomogeneous lattices in both two and three dimensions. In particular, it will be interesting to explore the analogue of Eq.~\ref{eq:loc_fin} in conformal lattices, e.g. flux line lattices under magnetic field gradients~\cite{menezes2019self,silva2020formation}, where the inhomogeneity can act not only through local dilation or contraction but also through local rotation. It is also widely known that density-mismatched bulk colloidal crystals in three dimensions exhibit smaller lattice constants at the bottom~\cite{russel1991colloidal,rutgers1996measurement,jensen2013rapid}, as the colloids further below become increasingly more crushed by the weight of the colloids and solvent above. 
It would be interesting to investigate the normal modes of these systems and see whether eigenfunction localization domains again appear at short wavelengths. Finally, in systems where the membrane is free-standing and out-of-plane fluctuations are important~\cite{nelson2004statistical}, it may be worthwhile to investigate how in-plane inhomogeneity affects the flexural phonons. 

% Note that 

\begin{acknowledgements}
We are grateful for helpful conversations with F. Spaepen and H. Zhou. G.H.Z. acknowledges support by the National Science Foundation Graduate Research Fellowship under Grant No. DGE1745303. This work was also supported by the NSF through the Harvard Materials Science and Engineering Center, via Grant No. DMR-2011754, as well as by Grant No. DMR-1608501.
\end{acknowledgements}

\appendix

\section{Dislocation pileups \label{app:pileup}}

% To obtain the vibrational spectrum, we add a kinetic term $\frac{m}{2} \sum_i \dot x_i^2$, where $m$ denotes some effective mass of the dislocation, to the static Hamiltonian $H$. (Note that in general, there are multiple types of kinetic terms that one could add, each leading to a different set of dynamics with the same equilibrium state, e.g. $\sum_i \dot x_i$ for damped dynamics. Our choice here for $\frac{m}{2} \dot x^2$ is arbitrary and does not affect the message of this section.) 
The dynamical Hamiltonian for a one-dimensional dislocation pileup embedded in a two-dimensional host crystal is given by~\cite{hirth1983theory},
\begin{eqnarray} \label{eq:Ham_discrete}
\mathcal{H} \left[ \left\{ x_{ n} \right\} \right] =&& \sum_{n=1}^N \frac{m}{2} \dot x_n^2 + \sum_ { n = 1 } ^ { N } b(x_n) \sigma_0  U(x_n) \notag \\
&&- \frac { Y} { 8 \pi  } \sum _ { n \neq m } b(x_n) b(x_m)  \ln \left| x_{ n } - x_{ m } \right|,
\end{eqnarray}
where $Y$ is the 2d Young's modulus controlling the repulsive pair potential, $b(x) = \pm b$ is the Burgers vector, $m$ is the effective mass of the dislocation, and $N$ is the total number of dislocations in the pileup. In the second term, ${\sigma_0 U(x) = \int_{-L/2}^x dx' \sigma(x')}$ comes from the Peach-Koehler force due to the applied shear stress $\sigma(x)$, where $\sigma_0$ measures the strength of the shear stress and $U(x)$, with dimensions of length, is the spatial profile of the potential experienced by the dislocations due to the shear stress. 

\subsection{Numerical determination of dislocation sites}

\begin{figure}[h]
\centering
\includegraphics[width=0.99\columnwidth]{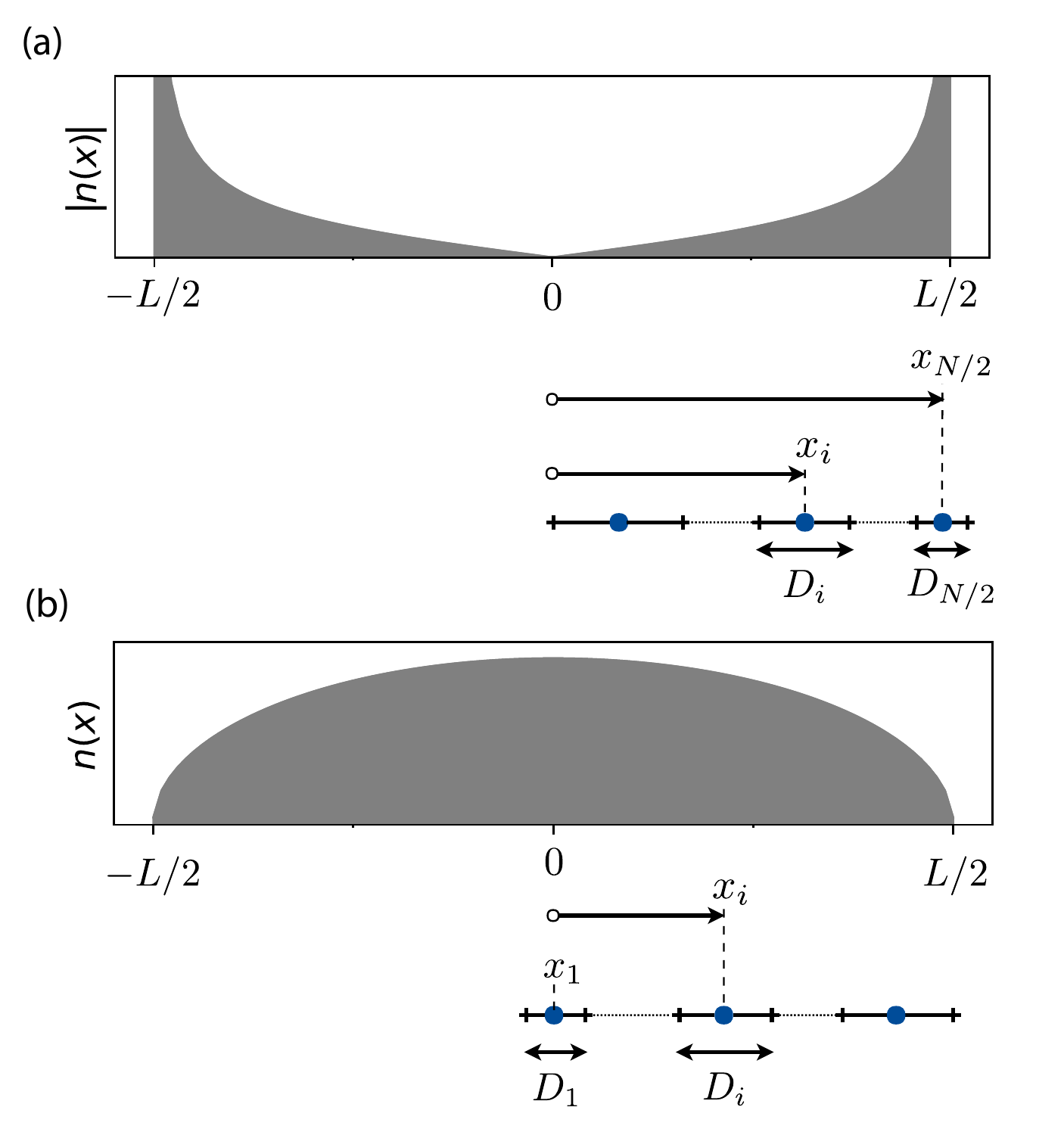}%
\caption{Schematic illustrating the numerical determination of dislocation positions for the double pileup (a) and the semicircle pileup (b). Gray shadings plot the absolute value of the dislocation charge densities $n_{\mathrm{D}}(x)$ and $n_{\mathrm{SC}}(x)$ in Eqs.~(\ref{eq:density_D}-\ref{eq:density_SC}). $x_i$ indicates the position of the $i$-th dislocation site from the origin, and $a_i$ indicates the lattice constant corresponding to the $i$-th dislocation site.\label{fig:lattice_sch}}
\end{figure}

The pileups in Fig.~\ref{fig:Vfull}a and b of the main text, referred to respectively as the double pileup and the semicircle pileup, each experience the following forms of shear stress,
\begin{eqnarray}
\sigma_\text{D}(x) &=& \sigma_0 \\
\sigma_\text{SC}(x) &=& \sigma_0 \frac{x}{L/2},
\end{eqnarray}
where $L$ is the length of the pileup, and the average dislocation charge densities are~\cite{hirth1983theory,zhang2020pileups}, as plotted in Fig.~\ref{fig:lattice_sch}. 
\begin{eqnarray}
n_{\mathrm{D}}(x)  &=\frac{4 \sigma_0}{Yb} \frac{x}{\sqrt{\left(\frac{L}{2}\right)^{2}-x^{2}}} \label{eq:density_D}\\
n_{\mathrm{SC}}(x) &=\frac{4 \sigma_0}{Yb} \sqrt{1-\left(\frac{x}{L / 2}\right)^{2}}. \label{eq:density_SC}
\end{eqnarray}

Note that for the double pileup, the sign of the defect charge changes across the origin. Additionally, although $n_D(x)$ is singular near $x = \pm L/2$, these defect densities vary smoothly elsewhere.

We now extract the equilibrium lattice positions numerically from the charge densities in Eqs.~(\ref{eq:density_D}-\ref{eq:density_SC}). Since the absolute value of both distributions are symmetric about the pileup centers, we compute the dislocation sites on the right half of the dislocation pileup (positive $x$ axis) and reflect across the vertical axis at $x=0$ (see Fig.~\ref{fig:lattice_sch}). 

On writing the dislocation positions as,
\begin{eqnarray}
x_i &=& x_{i - 1} + \frac{1}{2} \left[ D(x_{i-1})+ D(x_i) \right] \label{rec1} \\ %\frac{\zeta}{\sqrt{1 - \left(\frac{x}{L/2} \right)^2}} 
\end{eqnarray}
where $D(x_i) = |n(x_i)|^{-1}$ is the lattice constant at the position $x_i$ of the $i$-th dislocation, and rescaling spatially by half of the total length of the dislocation pileup, we have
\begin{eqnarray}
\bar x_i &=& \bar x_{i - 1} + \frac{1}{2} \left[ \bar D(\bar x_{i-1})+ \bar D(\bar x_i) \right]
\end{eqnarray} 
where $\bar x = x/(L/2), \bar D = D/(L/2)$. 
Since the dislocation density vanishes at the center for the double pileup and at the edge for the semicircle pileup, we solve for the dislocation positions iteratively starting from the center for the semicircle pileup and the edge for the double pileup, where the charge densities are finite.  
Upon using Eqs.~(\ref{eq:density_D}-\ref{eq:density_SC}) and defining $f_i \equiv \bar x_{i} + \frac{1}{2} \bar D(\bar x_i)$ and $g_i\equiv \bar x_i - \frac{1}{2} \bar D(\bar x_i)$, the position $\bar x_i$ of the $i$-th dislocation can be computed from the position of the $(i-1)$-th (or $(i+1)$-th) dislocation by numerically solving for the root of these quartic equations, starting from their respective boundary conditions, 
\begin{widetext}
\begin{gather}
\bar { x } _ { i ,D } ^ { 4 } - 2 g(\bar x_{i+1, D}) \bar { x } _ { i ,D } ^ { 3 } +  \left[ \left(\frac { Yb } { 4 L \sigma_0 }\right)^2 + g(\bar x_{i+1,D}) \right] \bar { x } _ { i ,D } ^ { 2 }- \left(\frac { Yb } { 4 L \sigma_0 }\right)^2 = 0, \quad \bar { x } _ { N/2,D } = 1 \\
\bar x_{i, SC}^4 - 2 f_{i-1} \bar x_{i, SC}^3 - \left( 1 - f(\bar x_{i-1, SC})^2 \right) \bar x_i^2 - 2 f(\bar x_{i-1,SC}) \bar x_i - f(\bar x_{i-1,SC})^2 - \frac{Yb}{8L\sigma_0} = 0, \quad \bar x_{1,SC} = 0. 
\end{gather}
\end{widetext}
\normalsize

\subsection{Dynamical matrix and real space eigenfunctions}

On decomposing the dislocation position as $x_n = R_n + u_n$ (where $R_n$ is the equilibrium location and $u_n$ is the displacement from the equilibrium lattice site), and assuming the displacements to be small relative to the lattice spacing at low temperatures, we expand the Hamiltonian up to quadratic order in $u_n$ as,
\begin{eqnarray}
\mathcal{H} \left[ \left\{ u _ { n} \right\} \right] &=& \sum_{n=1}^N \frac{m}{2} \dot u_n^2 + \sum_{n,m} M_{nm} u_n u_m,
\end{eqnarray}
% \begin{eqnarray}
% H_{U} \left[ \left\{ x _ { i } \right\} \right] &=& \frac{\sigma_0 b L}{2} \left[ \sum_i \frac{3 u_i^2 }{( 1 - R_i^2)} + \sum_{i \neq j} \frac { 1 } {\zeta \pi L}  \left( \frac { u _ { ij } } { R _ { i j} } \right) ^ { 2 } \right] \\
% H_{SC} \left[ \left\{ x _ { i } \right\} \right] &=& \frac{\sigma_0 b L}{4} \left[ \sum_i u_i^2 + \sum_{i \neq j} \frac { 2  } {\zeta \pi L } \left( \frac { 1 } { R _ { i j} } \right) ^ { 2 } u _ { ij }^2 \right] \\
% H_{D} \left[ \left\{ x _ { i } \right\} \right] &=& \frac { \sigma_0 b L } { 2 } \sum_{i \neq j}  \frac { 1 } { \zeta L \pi} \text{sign}(R_i R_j) \left( \frac { 1 } { R _ { i j} } \right) ^ { 2 } u _ { ij }^2
% \end{eqnarray} \normalsize %\normalsize
% The Hamiltonian can be expressed in matrix notation as
% \begin{equation}
% H_{U/SC/D} =  \sum_{i, j} \frac{1}{2} M^{U/SC/D}_{ij} u_i u_j,
% \end{equation}
where the diagonal $M_{n,n}$ and off-diagonal $M_{n, m \neq n}$ matrix elements for the double pileup are given by,
\begin{eqnarray}
 M^{D}_ { n,  n } & =&   \frac{\sigma_0 b}{ N} \sum _ { m \neq n } \frac { \text{sign}(R_n R_m) } { (R_n - R_m)^ { 2 } } \label{eq:M_first}\\ 
 M^{D}_ { n, m \neq n} & =&  - \frac{\sigma_0 b}{N} \frac { \text{sign}(R_n R_m) } { (R_n - R_m) ^ { 2 } },
\end{eqnarray}
and those of the semicircle pileup by,
\begin{eqnarray}  \label{eq:Ms}
%  M^{U}_ { i i } & =&  {\sigma_0 b} \left( \frac{3}{( 1 - R_i^2)} + \frac{1}{N} \sum _ { j \neq i } \frac { 1 } { R _ { i j } ^ { 2 } } \right) \\ 
%  M^{U} _ { i j } & =& - \frac{\sigma_0 b}{ N} \frac { 1 } { R _ { ij} ^ { 2 } } \\
 M^{SC} _ { n,n } & =&  {\sigma_0 b} \left( 1 + \frac{1}{N} \sum _ { m \neq n } \frac { 1 } { (R_n - R_m)^ { 2 } } \right) \\ 
 M^{SC}_ { n, m \neq n } & =&  - \frac{\sigma_0 b}{N } \frac { 1 } { (R_n - R_m)^ { 2 } }, \label{eq:M_last}
\end{eqnarray}
where the pileup lengths have been normalized such that $ x \in (-1, 1)$. Normalization conditions can be found in Table I of Ref.~\cite{zhang2020pileups}. 
Upon parameterizing $u _ { n} ( t ) = u _ { n } e ^ { -i \omega t }$, we obtain the equations of motion as,
\begin{eqnarray}
\Omega { u } _ { n }  = \sum_m M_{nm} u_m,
\end{eqnarray}
where $\Omega \equiv m \omega^2$ is the eigenenergy. 

As mentioned above, to obtain the matrix elements in Eqs.~(\ref{eq:M_first}-\ref{eq:M_last}), we compute the equilibrium dislocation positions $\{R_n \}$ by placing a dislocation at the center of the pileup $x=0$ and using Eqs.~(\ref{eq:density_D}) and (\ref{eq:density_SC}) to solve iteratively for the positions of its neighboring dislocations until the edge is reached. 
We then numerically compute the normal modes by solving the following eigenproblem,
\begin{eqnarray} \label{eq:eig}
\mathbf { M } \mathbf { u }^ { (\alpha )} = \Omega _ { \alpha } \mathbf { u }^{ (\alpha) },
\end{eqnarray}
where $\bf{M}$ is the dynamical matrix with elements $M_{n,m}$, and $\Omega_\alpha \equiv m \omega_\alpha^2$ and $\bf{u}^{(\alpha)}$ are the eigenenergies and eigenfunctions of the $\alpha$-th eigenmode. 
% Note that the eigenfrequency has dimensions $[\Omega] = \frac{E}{L^2}$. 

\subsection{Phonon spectrum and band edge}

The full dispersion relation for a uniform one-dimensional pileup is given by~\cite{zhang2020pileups},
\begin{eqnarray} \label{eq:pileup_full_disp}
    \Omega(q) &=& \frac{1}{D^2} \frac{Yb^2}{4 \pi} \sum_{n=1}^{\infty} \frac{1-\cos (n q D)}{n^{2}}  \\
    &=&\frac{1}{D^2} \frac{Yb^2}{8 \pi} \left \{ \frac{\pi^2}{3} - \left[\text{Li}_2 (e^{-i q D}) + \text{Li}_2 (e^{i q D}) \right] \right \},  
\end{eqnarray}
where $\text{Li}_n(z)$ is the polylogarithmic function (or Jonquière's function)~\cite{erdelyi1953higher}, $D$ is the the average dislocation spacing, $Y$ is the Young's modulus of the 2d host crystal, and $b$ is the magnitude of the Burgers vector. 
The band edge, which occurs at the boundary of the first Brillouin zone $q= \pi/D$, is then obtained from Eq.~(\ref{eq:pileup_BE}) according to Eq.~(\ref{eq:pileup_full_disp}).

We show the evolution of the localization domains for the double pileup and the semicircle pileup in Fig.~\ref{fig:Vfull}. According to Eqs.~(\ref{eq:tp_all_springs}), (\ref{eq:density_D}), and (\ref{eq:density_SC}), the localization domains $(x_\alpha^-, x^+_\alpha)$ for these pileups satisfy,
\begin{eqnarray} \label{eq:omega_pileup}
\Omega_\alpha = \frac{\pi \sigma_0^2}{Y} \times \begin{cases} \frac{x_\alpha^2}{\left( \frac{L}{2} \right)^2 - x_\alpha^2 }, \quad \quad ~ \text{Double pileup}\\
1 - \left(\frac{x_\alpha}{L/2} \right)^2, \quad \text{Semicircle}\end{cases}
\end{eqnarray}
Note that $Y$ and $\sigma_0$ have units of $E/L^2$ in 1d, so the dimensions on both sides of the equation match. As shown in Fig.~\ref{fig:Vfull}, the band edge shown in Eq.~(\ref{eq:omega_pileup}) agrees with the numerical eigenfunctions. 
As the energy of the excitation mode increases, the eigenfunction becomes increasingly more localized towards the densest parts of the lattice. For the semicircle pileup, this corresponds to the middle of the pileup, while for the double pileup, this is the region near the pileup edges.

\section{1d chain of balls and springs under gravity \label{app:1d}}

An instructive toy model is a one-dimensional chain of balls of mass $m$, connected by springs with stiffness $k_s$, subject to gravity on a plane inclined by angle $\theta$ with respect to the ground. Although that the gravitational force causes the density of the masses to be inhomogeneous along the inclined plane, the equations of motion and thus the phonons remain identical to that of the homogeneous lattice without the effects of gravity. Thus, for this purely harmonic problem, all effects of the lattice inhomogeneity can be redefined away. 

\begin{figure}
    \centering
    \includegraphics[width=0.95\columnwidth]{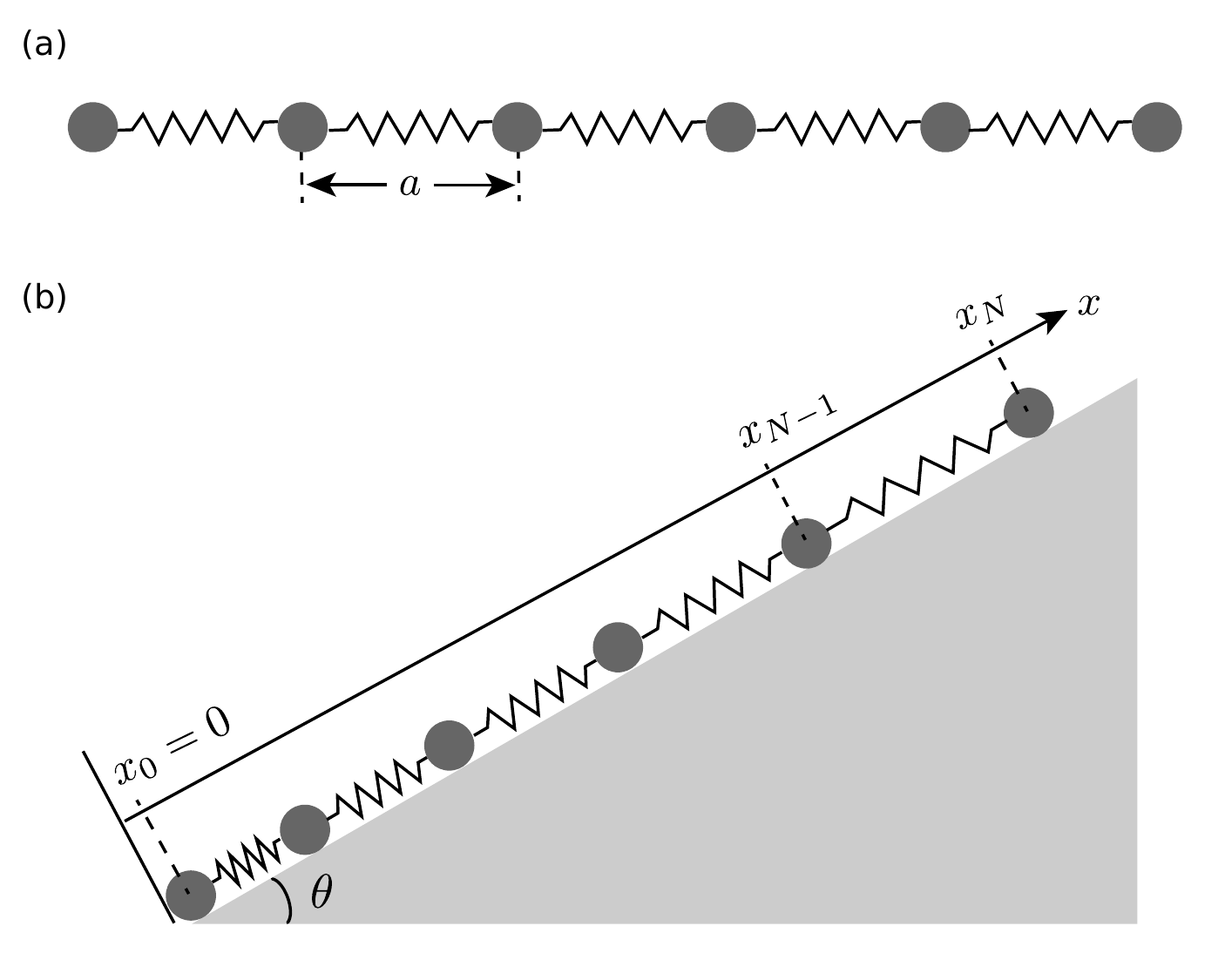}
    \caption{Schematic for 1d problem of springs and masses on a flat surface (a) and on an inclined plane of angle $\theta$ (b). $a$ is the equilibrium distance between two neighboring balls on a flat plane. $x_n$ denotes the position of the $n$-th ball along the inclined plane from the bottommost ball at $x_0=0$.}
    \label{fig:1d_schm}
\end{figure}

The energy that leads to phonon dynamics in this system is 
\begin{eqnarray} 
H[\{x_n \}] &=& \sum_{n=1}^{N} \left [  \frac{m}{2} \dot x_n^2 + \frac{1}{2} k_s \left(x_{n+1}-x_{n}-a\right)^{2}-m^{\prime} g  x_{n} \right],  \notag \\ \label{eq:E_sp_g}
\end{eqnarray}
where $m^{\prime}=m \sin \theta$,  $g$ is the gravitational acceleration, and $x_n$ is the position (along the inclined plane) of the $n$-th ball. The lowest point of the inclined plane is met by a wall such that the first ball (indexed by $n = 0$) has fixed position centered at $x_0 = 0$ (see Fig.~\ref{fig:1d_schm}). 
% Upon expanding the first term in Eq.~\ref{eq:E_sp_g}, we have
% \begin{eqnarray}
% E\left(x_{1}, \ldots, x_{n}\right) &=&\frac{1}{2} k \sum_{n}\left[x_{n+1}^{2}-2 x_{n+1} x_{n}+x_{n}^{2} \right.\\
% &&\left.  -2 a\left(x_{n+1}-x_{n}\right)+a^{2}\right] - m' g \sum_n x_n.  \notag
% \end{eqnarray}
% To calculate the equilibrium positions $x_n^*$, we find the minimum of the energy with respect to the ball positions
% \begin{eqnarray}
% \frac{\partial E}{\partial x_{n}} &=& \frac{1}{2} k\left[2 x_{n}-2 x_{n+1}+2 a+2 x_{n}-2 x_{n-1}-2 a \right] - m' g \notag \\
% \\
% &=& k\left[2 x_{n}-x_{n+1}-x_{n-1} \quad\right]-m' g. 
% \end{eqnarray}
% The equilibrium condition requires $\partial E (x_n = x_n^*)/ \partial x_n =0 ~\forall n$,
% \begin{eqnarray} \label{eq:fb_homo_1d}
% \frac{d E}{d x_{n}}=0=k  \left[ 2 x^*_{n}-x^*_{n+1}-x^*_{n-1} \right] - m g^{\prime}
% \end{eqnarray}
% We use the ansatz
% \begin{eqnarray}
% x^*_n = na + u^*_n,
% \end{eqnarray}
% where $na$ is the lattice position without gravitational force. 
% Eq.~\ref{eq:fb_homo_1d} then becomes
% \begin{eqnarray}
% \frac{\partial E}{d x_{n}}&=&k[2 n-n-1-n+1] a + k[ 2 u_n^* - u_{n+1}^* - u_{n-1}^* ] -m' g\\
% 0 &=& k[ 2 u_n^* - u_{n+1}^* - u_{n-1}^* ] -m' g \\
% &=& k \frac{\partial^2 u_n^*}{\partial n^2} - m' g \label{eq:fb_1d}
% \end{eqnarray}
% Solving the second order ordinary differential equation then gives us
% \begin{eqnarray} \label{eq:unstar}
% u_n^* = \frac{m'g}{2 k} n^2 + n \alpha + \beta,
% \end{eqnarray}
% where $\alpha$ and $\beta$ are constants determined by boundary conditions, 
Upon writing the new equilibrium position as $x_n^* = na + u_n^*$, minimizing the energy, and applying the following boundary conditions (zero displacement at $n=0$ and zero stress at $n=N$),
\begin{eqnarray} \label{eq:bc}
u^*(n=0) = 0, \quad \frac{\partial u^*(n)}{\partial n} \Big|_N = 0. 
\end{eqnarray}
we obtain the equilibrium lattice positions $\{x_n^* \}$ as,
% :
% Upon plugging in Eq.~\ref{eq:unstar} into Eq.~\ref{eq:bc}, we have
% \begin{eqnarray}
% \beta = \frac{m'g}{k}\left( N - \frac{1}{2} \right), \quad \alpha = - \frac{m' g}{k} N
% \end{eqnarray}
% and arrive at the following equilibrium lattice positions
\begin{eqnarray}
x^*(n) &=& na + \frac{m^{\prime} g}{2 k_s}\left[n^{2}-2 N n\right].
\end{eqnarray}

% We now see if the inhomogeneous equilibrium positions affect the vibrational spectrum. 
% Upon incorporate a dynamical kinetic term, our Hamiltonian is
% \begin{eqnarray}
% H[\{x_n \}] &=& \sum_{n} \frac{m}{2} \dot u_n^2 + \sum_n \left[ \frac{k}{2}  \left( x_{n+1}^* + u_{n+1} - x_{n}^* - u_{n} \right)^2 \right. \notag\\
% && \left. - m' g \left( x_n^* + u_n \right) \right]  
% \end{eqnarray}
Upon decomposing the mass position in Eq.~(\ref{eq:E_sp_g}) as ${x_n = x_n^* + u_n}$, where $u_n$ is the displacement away from the equilibrium position $x_n^*$, and noting that terms linear in displacement vanish due to force balance, we are left with,
\begin{eqnarray} \label{eq:H_1d_after}
H[\{x_n \}] &=& \sum_{n} \frac{m}{2} \dot u_n^2 + \sum_n \frac{k_s}{2}  \left( u_{n+1}^2 +  u_{n}^2 - 2u_{n+1}u_n \right)  \notag \\
&&  - \sum_n m' g x_n^*. 
\end{eqnarray}
% and thus the Lagrangian
% \begin{eqnarray}
% \mathcal{L}[\{x_n \}] &=& \sum_{n} \frac{m}{2} \dot u_n^2 - \sum_n \frac{k}{2}  \left( u_{n+1}^2 +  u_{n}^2 - 2u_{n+1}u_n \right) \notag \\
% &&+ \sum_n m' g x_n^*, 
% \end{eqnarray}
Since the gravitational term in Eq.~(\ref{eq:H_1d_after}) only contributes a constant shift to the energy, it does not appear in the equations of motions, 
% The equation of motions 
% \begin{eqnarray}
% \frac{ \partial \mathcal{L}}{\partial u_n} - \frac{\partial}{\partial t} \frac{ \partial \mathcal{L}}{\partial \dot u_n} = 0
% \end{eqnarray}
\begin{eqnarray} \label{eq:eom_homo}
- m \ddot u_n = k_s \left( 2 u_n - u_{n+1} - u_{n-1} \right),
\end{eqnarray}
which are identical to the equations of motion for a homogeneous chain of balls and springs. 
The phonon spectrum, upon assuming wave-like solutions $u_n = \frac{1}{\sqrt{N}} e^{i q n a} e^{i \omega t}$, is also unchanged,
% Thus, the spectrum $\omega^2$ is unaffected by the inhomogeneous equilibrium positions. We can see from Eq.~\ref{eq:eom_homo}, however, that the equations of motion would change if the spring constant were to become inhomogeneous $k \rightarrow k_n$. 
% Inserting this into Eq.~\ref{eq:eom_homo} gives us
% \begin{eqnarray}
% m \omega^2 = 2 k( 1 -  \cos q a ). 
% \end{eqnarray}
% The dispersion relation for a homogeneous spring is thus given by
\begin{eqnarray}
\omega^2(q) = 4\frac{k_s}{m}  \sin^2 \left( \frac{ qa}{2}\right) . 
\end{eqnarray}
Note that the band edge at $q = \pi/a$ is independent of the lattice spacing $a$, so the system does not exhibit the quasi-localization phenomena described in the main body of this paper. In summary, the inhomogeneity in this toy problem does not affect the phonon spectrum, because it does not appear in the interaction strengths. However, if the stiffness $k_s$ of the spring connecting two masses were to depend on the distance between them, then $k_s \rightarrow k_s(x^*_n)$ in the presence of inhomogeneity and that \textit{could} change the equations of motions in interesting ways.   
% Upon replacing $q = \frac{2 \pi}{a}$, we see that the dispersion relation becomes independent of $a$, and thus its normal modes does not exhibit the localization phenomena seen in later sections, where the spring constants $k$ becomes inhomogeneous. 
\section{Elastic deformation of a flat elastic sheet subjected to gravitational pressure in the plane of the sheet} \label{app:sheet}

\begin{figure}
    \centering
    \includegraphics[width=1\columnwidth]{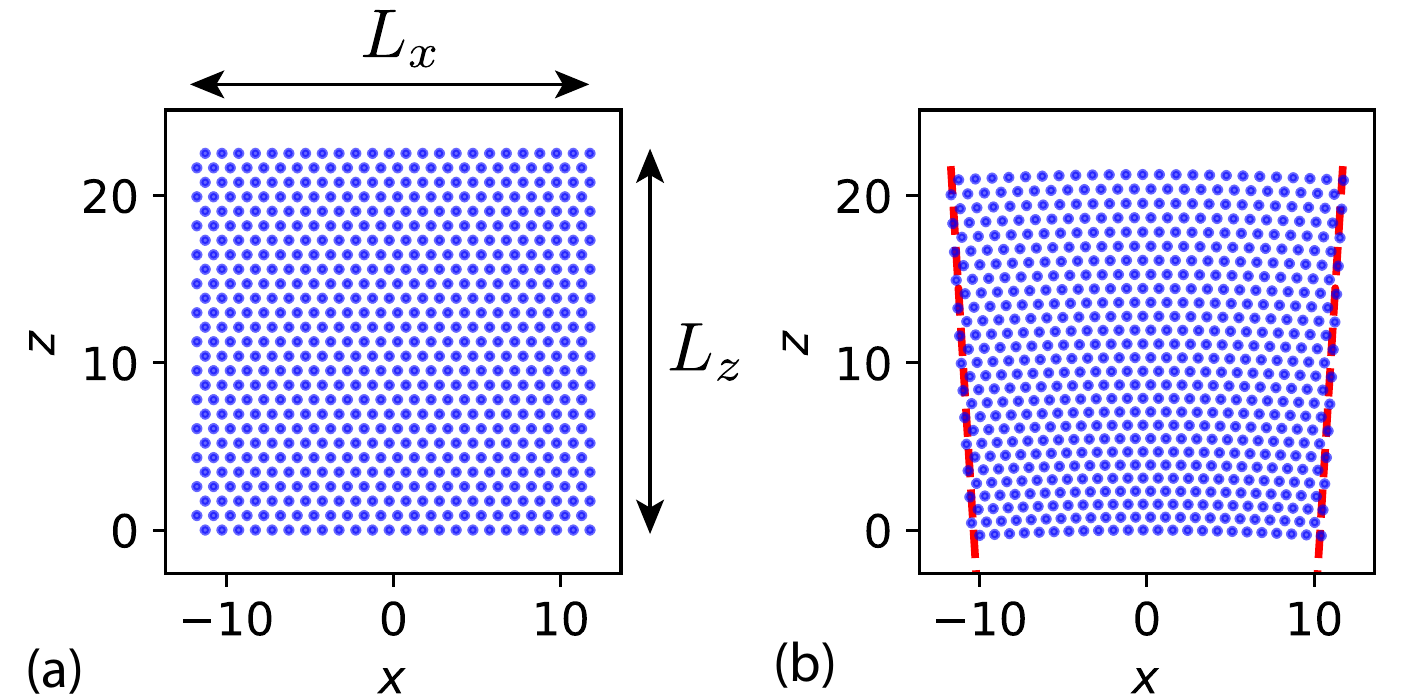}
    \caption{(a): The unfrustrated ground state configuration of $N$ particles interacting via the Lennard-Jones potential in the flat plane is the triangular lattice with lattice constant $a_0$, height $L_z$, and width $L_x$. (b): When experiencing an isotropic gravitational pressure, the flat sheet becomes geometrically frustrated, curving the upper and lower boundaries and rendering solutions from simple linear elasticity theory invalid. As discussed in the main text, this frustration can be avoided by wrapping the top and bottom boundaries around to form a truncated cone.}
    \label{fig:sheet}
\end{figure}
We examine a 2d sheet in the flat plane with uniform elastic coupling constants under gravitational pressure using continuum elasticity theory, and show that the gravitational potential frustrates the ground state when the particles are constrained to remain on a flat plane parallel to the gravitational force. In contrast, as shown in the main text, when we allow the sheet of particles to escape into the third dimension by imposing cylindrical boundary conditions, the frustration can be gauged away, i.e. eliminated by a change of variables.  

The free energy of a two-dimensional sheet in the {$x$-$z$} plane under an isotropic pressure that increases with decreasing vertical coordinate $z$ is given by,
\begin{eqnarray} \label{eq:F_grav_app}
F = \frac{1}{2} \int d^2 r  \left( 2 \mu u_{ik}^2 + \lambda u_{ii}^2  + 2 \alpha_0 (L_z-z) u_{ii} \right),
\end{eqnarray}
where $\delta p = \alpha_0 (L-z)$ is an isotropic $z$-dependent pressure, $u_{ij} = \frac{1}{2} (\frac{\partial u_i}{\partial r_j} + \frac{\partial u_j}{\partial r_i} ) $ is the strain tensor, $\mu$ and $\lambda$ are the first and second 2d Lam\'{e} coefficients. Here $\alpha_0$ is proportional to the gravitational constant $g$ (see Eq.~(\ref{eq:alpha0_g})) and $\vec u(\vec r)$ is the displacement of the sheet from its  $\alpha_0 = 0$ equilibrium configuration at position $\vec r$.  
The non-deformed ($\alpha_0 = 0$) sheet extends from $[0, L_z]$ along $z$ and from $[-L_x/2, L_x/2]$ along $x$ (Fig.~\ref{fig:sheet}a). The natural boundary conditions for this problem are anchoring the flat bottom edge of the sheet at the base, which requires that the vertical displacements $u_z$ vanish at $z = 0$ and the stresses vanish at $z = L_z$. 

We now find the set of strains $\bar u_{ij}$ that minimize Eq.~(\ref{eq:F_grav_app}). First, since the gravitational term is isotropic and does not depend on $u_{zx}$, we immediately have
\begin{eqnarray} \label{eq:uxy}
\bar u_{zx} = 0. 
\end{eqnarray}
Next, upon setting the derivatives of the free energy $F$ with respect to $u_{zz}$ and $u_{xx}$ to zero as,
\begin{eqnarray} \label{eq:dFduii}
0&=&  (2 \mu + \lambda) \bar u_{zz} (r) + \bar u_{xx} (r)+ \alpha_0(L_z -z) \\
0 &=&(2 \mu + \lambda)  \bar u_{xx} (r) + \bar u_{zz} (r)+ \alpha_0(L_z -z)  
\end{eqnarray}
we obtain
\begin{eqnarray} \label{eq:uxx}
\bar u_{zz} = \bar u_{xx} = \frac{\alpha_0 (z - L_z)}{2B},
\end{eqnarray}
where $B = \mu + \lambda$ is the 2d bulk modulus of the sheet. 

Since $\bar u_{zz}=\bar u_{xx}=0$ at $z = L_z$, the stresses $\sigma_{zz}$ and $\sigma_{xx}$ also vanish there, so the free boundary condition at $z = L_z $ is satisfied. 
Upon directly integrating Eq.~(\ref{eq:uxx}), we obtain,
\begin{eqnarray} 
\bar u_z &=& \frac{\alpha_0}{2 B} \left( \frac{z^2}{2} - L_z z \right) + \phi(x)  \label{eq:u_phi}\\
\bar u_x &=& \frac{\alpha_0}{2 B} (z - L_z) x + \xi(z).  \label{eq:u_xi}
\end{eqnarray}
To determine $\phi(x)$ and $\xi(z)$, we use Eq.~(\ref{eq:uxy}),
\begin{eqnarray}
\bar u_{zx} = 0 = \frac{d \phi(x)}{dx} + \frac{\alpha_0}{2 B} x + \frac{d \xi(z)}{dz}
\end{eqnarray}
which leads to,
\begin{eqnarray}
\xi(z) &=& c \\
\phi(x) &=& - \frac{\alpha_0}{2 B}  \frac{x^2}{2} + d,
\end{eqnarray}
where $c$ and $d$ are constants, and Eqs.~(\ref{eq:u_phi}) and (\ref{eq:u_xi}) become,
\begin{eqnarray} 
\bar u_z &=& \frac{\alpha_0}{2 B} \left( \frac{z^2}{2} - L_z z - \frac{x^2}{2}\right) + d \label{eq:uz_frus}\\
\bar u_x &=& \frac{\alpha_0}{2 B} (z - L_z) x + c. \label{eq:ux_frus}
\end{eqnarray}
However, we immediately see that the fixed displacement boundary condition at the bottom edge of a planar sheet (as would be the case if it rested on a flat surface) cannot be satisfied, as there is no choice of $c$ and $d$ in Eqs.~(\ref{eq:uz_frus}) and (\ref{eq:ux_frus}) such that $\bar u_z  (x, z = 0) =\bar u_x (x, z = 0) = 0$ for all $x \in [-L_x/2, L_x/2]$. 
% \begin{eqnarray}
% \bar u_z(z = 0) &=& \frac{\alpha_0}{2 B} \left( - \frac{x^2}{2}\right) + d \neq 0 \label{eq:uz_frus}\\
% \bar u_x(z =0) &=& \frac{\alpha_0}{2 B} (- L_z x)  + c \neq 0 \label{eq:ux_frus}
% \end{eqnarray}

As an approximation, we can set $c = d = 0$ in Eqs.~(\ref{eq:uz_frus}) and (\ref{eq:ux_frus}) such that only the $(x,z) = (0,0)$ point has zero displacement. Fig.~\ref{fig:sheet}b then plots the triangular lattice sites displaced according to
\begin{eqnarray} 
\bar u_z &=& \frac{\alpha_0}{2 B} \left( \frac{z^2}{2} - L_z z - \frac{x^2}{2}\right) \label{eq:solx_app} \\
\bar u_x &=& \frac{\alpha_0}{2 B} (z - L_z) x. \label{eq:soly_app}
\end{eqnarray}
The curl of the 2d displacement field $(u_x, u_z)$ shows that the bond angles $\vartheta$ between nearest neighbors on the lattice are in fact rotating as a function $x$,
\begin{eqnarray}
\vartheta(x)= \frac{1}{2} \left( \frac{\partial u_z}{\partial x} -  \frac{\partial u_x}{\partial z} \right) = - \frac{\alpha_0  }{2 B} x. 
\end{eqnarray}

Since Eqs. (\ref{eq:solx_app}) and (\ref{eq:soly_app}) cannot satisfy the bottom boundary conditions, the elasticity solutions are not valid near the bottom of the sheet. This problem also arises in three dimensions for rods deforming under a gravitational field paralle to the rod axis, where planar boundary conditions at the bottom render linear elastic solutions invalid near the lower end of the rod~\cite{landau2012theory}.
Thus, for flat 2d sheets in the flat $x$-$z$ plane, there is no smooth deformation that can minimize the free energy in Eq.~(\ref{eq:F_grav_app}). In other words, we cannot ``gauge away'' the effect of gravitational pressure in the 2d plane. 
% , and hence provides additional strain energy compared to the ground state in the absence of the gravitational potential, for which $F = 0$. 
In this situation, the extra strain energy for this wedge-shaped sheet might possibly be reduced by introducing defects, such as a vertical grain boundary at $x=0$~\cite{zhang2020statistical,meng2021defects}. However, we show in the main text that, remarkably, this geometric frustration can be removed entirely by wrapping the horizontal direction $x$ around to form a cylinder and thus allowing the sheet to curve into the third dimension. 
\section{Displacements under hydrostatic compression in cone coordinates}

In this section, we derive the displacements in terms of cone coordinates $r$ (Eqs.~(\ref{eq:solr}-\ref{eq:solrad})) using the displacements in the Cartesian frame $(x,z)$. First, we show that the inner edge of an annulus on the polar plane corresponds to the bottom edge of our deformed sheet in the small cone angle limit. Next, we calculate the azimuthal displacement $\bar u_\phi$ using the polar cone angle and the circumference of the bottom edge of the truncated cone. Finally, upon requiring that the compression must be isotropic along both the azimuthal and axial directions, we extract the longitudinal displacement $\bar u_r$. 

Recall that the displacements in the flat plane under hydrostatic compression are given by Eqs.~(\ref{eq:solx}-\ref{eq:soly}) in the main text as,
\begin{eqnarray}
\bar u_z = \frac{\alpha_0}{2 B} \left( \frac{z^2}{2} - L_z z - \frac{x^2}{2} \right), \quad \bar u_x = \frac{\alpha_0}{2 B} \left( z - L_z \right) x, \label{eq:disp}
\end{eqnarray}
and the polar cone angle in the small angle limit is given by Eq.~(\ref{eq:apex}) as,
\begin{eqnarray}
\Phi = \frac{\alpha_0}{2 B} L_x + O\left( \left( \frac{\alpha_0}{2 B} L_z \right)^2 \right). \label{eq:phi_small}
\end{eqnarray}Note that this polar cone angle vanishes when the gravitational pressure $\sim \alpha_0 \rightarrow 0$ and when the material becomes incompressible $B = \mu + \lambda \rightarrow 0$. 

\begin{figure}
    \centering
    \includegraphics[width=0.9\columnwidth]{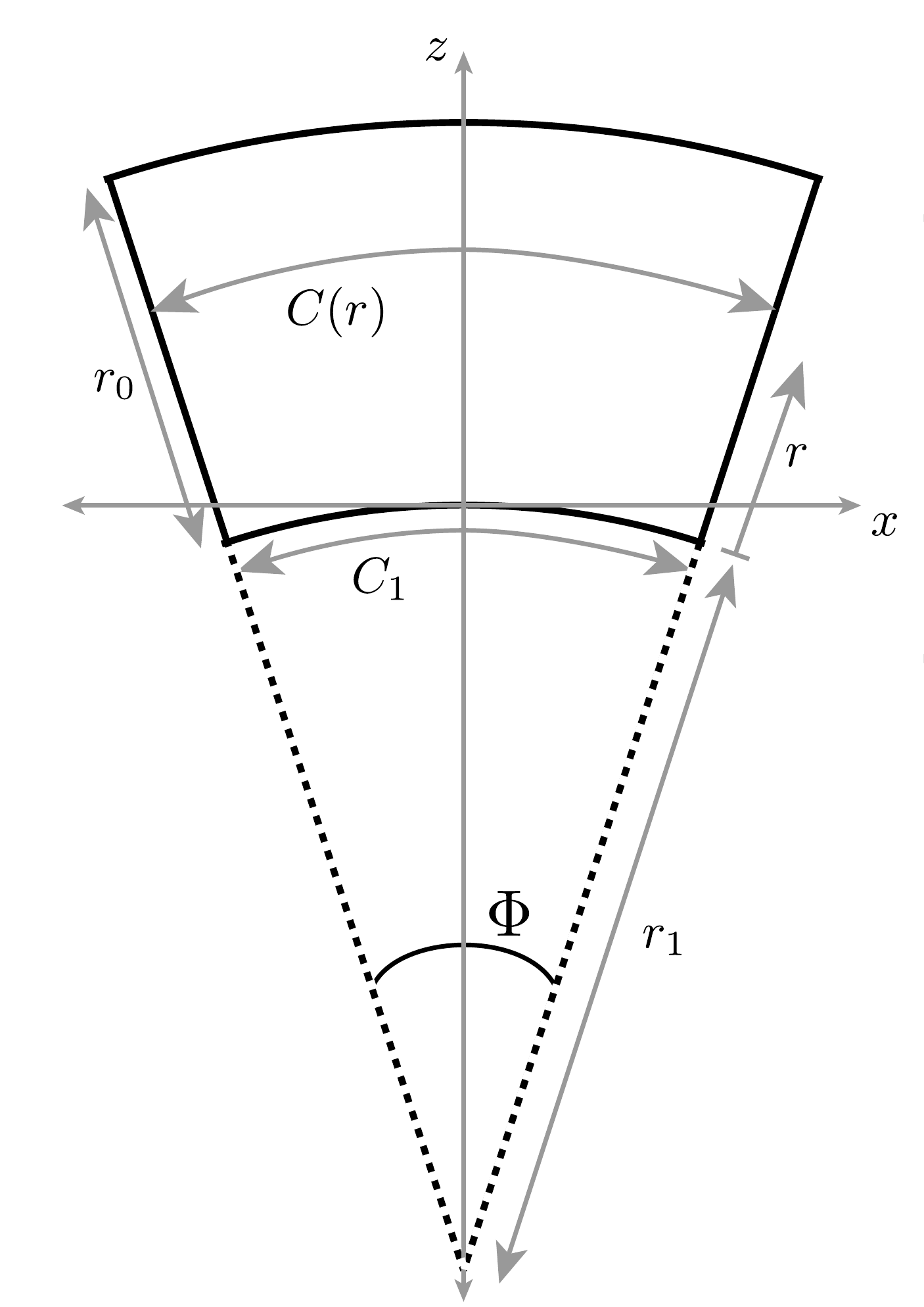}
    \caption{Schematic illustrating the quantities mentioned in text. The solid black lines bound a 2d planar crystal deformed in the $(x,z)$-plane by hydrostatic compression. Although the boundary conditions at the bottom can only be satisfied at the single point $(x,z)=(0,0)$ in the plane, this problem vanishes when we wrap this crystal around to form a small-angle truncated cone with polar angle $\Phi$. Here, $r$ is the distance up the flanks of the cone from its base.}
    \label{fig:schm}
\end{figure}

Upon letting $r_1$ denote the inner radius of the annulus (the polar projection of a truncated cone, see Fig.~\ref{fig:schm}), the inner annular rim is given by,
\begin{eqnarray}
x^2 + (z + r_1)^2 &=& r_1^2 
\end{eqnarray}
from which it follows that
\begin{eqnarray}
x^2 
&=& - 2 z r_1 \left( 1 + \frac{z}{2 r_1} \right) \label{eq:x2}  \\
&\approx& - 2 z r_1, \notag
\end{eqnarray}
where for small cone angles, $z \ll r_1$, so the second term can be neglected to lowest order. We calculate $r_1$ specifying the tip of the cone from
\begin{eqnarray}
\Phi r_1 = C_1 , \label{eq:r1def}
\end{eqnarray}
where $C_1$ is the circumference of the cross-section of the cone at $r=0$ (see Fig.~\ref{fig:schm}). Upon using Eqs.~(\ref{eq:disp}), $C_1$ is given approximately by,
\small
\begin{eqnarray}
\left( \frac{C_1}{2} \right)^2 & \approx & \left (\frac{L_x}{2} + \bar u_x (x = L_x/2, z=0) \right)^2 + \bar u_z(x = L_x/2, z=0)^2 \notag \\
% &=& \left(\frac{L_x}{2} \right)^2 \left[ 1 - 2 \frac{ \alpha_0}{2B} L_z  + \left( \frac{ \alpha_0}{2B}\right)^2 \left( L_z^2 + \frac{1}{4} \left(\frac{L_x}{2}  \right)^2 \right) \right] \\
&=&\left(\frac{L_x}{2} \right)^2 \left\{ 1 - 2\frac{ \alpha_0}{2B} L_z  + O\left[ \left(\frac{ \alpha_0}{2B} L_z \right)^2,  \left(\frac{ \alpha_0}{2B} L_x \right)^2 \right] \right \}, \notag \\
\end{eqnarray} \normalsize
which, to linear order in the cone angle, gives
\begin{eqnarray}
C_1 & \approx & L_x  \left( 1 - \frac{ \alpha_0}{2B} L_z \right).  \label{eq:C1res}
\end{eqnarray}
Upon substituting Eq.~(\ref{eq:C1res}) into Eq.~(\ref{eq:r1def}) and using Eq.~(\ref{eq:phi_small}), we then obtain
\begin{equation} \label{eq:r1}
r_1 = \frac{\left( 1 - \frac{ \alpha_0}{2B} L_z \right)}{\frac{\alpha_0}{2 B}}. 
\end{equation}
On inserting Eq.~(\ref{eq:r1}) into Eq.~(\ref{eq:x2}), we have
\begin{eqnarray}
z &=&- \frac{x^2}{2r_1} \notag\\
&=& - \frac{x^2}{2} \frac{\frac{\alpha_0}{2 B}}{\left( 1 - \frac{ \alpha_0}{2B} L_z \right)} \notag \\
&\approx & - \frac{x^2}{2} \frac{ \alpha_0}{2 B} + O \left(L_x \left(\frac{\alpha_0 L_x}{2 B} \right)\left(\frac{\alpha_0 L_z}{2 B} \right)\right),\label{eq:zcone} 
\end{eqnarray}
so Eq.~(\ref{eq:zcone}) is indeed equivalent to $\bar u_z(x, z=0)$ in Eq.~(\ref{eq:disp}),
\begin{eqnarray}
\bar u_z(x, z=0) = - \frac{x^2}{2} \frac{ \alpha_0}{2 B}. \label{eq:uzz0}
\end{eqnarray}

In the small cone angle limit where $r_1 \gg r$ and $r_1 \gg z$, we approximate the circumference $C(r)$ as a function of $r$ and the polar cone angle $\Phi$  as (see Fig.~\ref{fig:schm}), \small
\begin{eqnarray}
C(r) = \Phi(r_1 + r) = C_1 + \Phi r = L_x \left[ 1 + \frac{\alpha_0}{2B}(r - L_z) \right]. 
\end{eqnarray} \normalsize
We can then solve for the radial displacement $\bar u_R(r)$ as,
\begin{eqnarray}
2 \pi \bar u_R(r)  &=& C(r) - L_x  = \frac{\alpha_0}{2 B} (r - L_z)L_x 
\end{eqnarray}
which leads to
\begin{eqnarray}
\bar u_R(r)  &= & \frac{1}{2\pi}  \frac{\alpha_0}{2 B} (r - L_z)L_x.\label{eq:urad_fin}
\end{eqnarray}
Note that the radial displacement becomes more negative as one moves down the flanks of the cone by decreasing $r$. The radial displacement can be written as the usual azimuthal displacement $\bar u_\phi(r, \phi)$ around the axis of the cone on its tangent plane as,
\begin{eqnarray}
\bar u_\phi(r, \phi) = \phi \bar u_R(r). \label{eq:uphi_fin}
\end{eqnarray}

To obtain the other component of the displacement field, we note that since the gravitational pressure is isotropic, the compression factor has to be the same along both the axial ($r$) and azimuthal ($\phi$) directions:
\begin{eqnarray}
\frac{ d\bar u_r(R)}{dr} = \frac{ d\bar u_\phi(r, \phi)}{d(L_x \phi / 2 \pi)} = \frac{\alpha_0}{2 B} (r - L_z). 
\end{eqnarray}
Upon integrating and imposing cylindrical boundary conditions, the axial displacement $\bar u_r$ is given by
\begin{eqnarray}
\bar u_r (r) = \frac{ \alpha_0}{2 B} \left( \frac{r^2}{2} - L_z r \right). \label{eq:ur_fin}
\end{eqnarray}
Using the fact that $L_z$ is well approximated by the longitudinal cone length $r_0$ in Eqs.~(\ref{eq:uphi_fin}) and (\ref{eq:ur_fin}) to linear order in the cone angle, we obtain Eqs.~(\ref{eq:solr}-\ref{eq:solrad}) in the main text.

\section{Force balance for particles with nonlinear interaction potentials \label{app:force}}

In this section, we show that under a gravitational potential, the equilibrium displacements of a lattice with elastic coefficients $\mu(r), \lambda(r)$ that depend on the local lattice spacing is the same as that of a lattice with constant elastic coefficients $\mu, \lambda$, provided that the gravitational pressure is sufficiently week. This latter condition corresponds exactly to the small cone angle limit. 

To examine the nonlinear feedback between an inhomogeneous lattice and elastic constants that depend on the local lattice spacing, we write the local Lam\'e coefficients $\lambda (\vec r)$ and $\mu(\vec r)$ in terms of the strain tensor. 
We first write the $x$-th component of the lattice spacing $a_{x,n}$ of the $n$-th particle as,
\begin{eqnarray} \label{eq:a_fb}
a_{x, n} =  a_0 + (\bar u_{x, n+1_x} - \bar u_{x, n} ) \approx a_0 ( 1 + \bar u_{xx}),
\end{eqnarray}
where $a_0$ is the lattice constant of the undeformed sheet ($\alpha_0 = 0$). Here, $\bar u_{x, n}$ is the $x$-th component of the displacement of the $n$-th particle from its position in the unpressurized lattice, and $n+1_x$ denotes the index of its neighbor in the $\hat e_x$ direction. A similar expression follows for $a_{y,n}$.

To reduce notational clutter, we will show the calculation explicitly for interaction potentials of the power law form $V(\vec r) = A/|\vec r|^t$.  The results can then be transcribed to the Lennard-Jones potential, which is a sum of two power law terms with $t = 6$ and $12$. Upon using Eqs.~\ref{eq:eff_lame} and \ref{eq:uxx}, the space-dependent elastic coupling coefficients for Lennard-Jones particles on a cone may be written as,
\begin{eqnarray}
\lambda (\vec r) &=&  \frac{\lambda_0 }{ \left (1 + \frac{\alpha_0}{B_0} (z - L_z) \right)  ^{t+2}} 
\end{eqnarray}
\begin{eqnarray}
\mu (\vec r) &=&  \frac{\mu_0}{\left (1 + \frac{\alpha_0}{B_0} (z - L_z) \right)  ^{t+2}} 
\end{eqnarray}
where $B_0 = \lambda_0 + \mu_0$ and $\mu_0, \lambda_0$ are the elastic constants of the undeformed lattice. 

Having determined the form of the nonlinear elastic coupling coefficients $\lambda(\vec r)$  and $\mu(\vec r)$, we want to solve for the equilibrium positions $\{R_n\}$ of the particles under gravitational pressure by minimizing the following free energy,
\begin{eqnarray} \label{eq:F_grav_nonlinear}
F = \frac{1}{2} \int d^2 r  \left( 2 \mu (\vec r) u_{ij}^2 + \lambda(\vec r)  u_{ii}^2  + 2 \alpha_0 (L_z-z) u_{ii} \right). 
\end{eqnarray}
Taking derivatives of $F$ with respect to the components of the strain tensor $u_{ij}$ yields, upon using Eq.~\ref{eq:a_fb} for the local lattice spacing, 
\begin{widetext} \normalsize 
\begin{equation} \label{eq:fb_uyy} 
\begin{aligned}
0 &= (2 \mu(\vec r) + \lambda(\vec r) )\bar u_{xx} +  \lambda(\vec r) \bar u_{zz} + \alpha_0(L_z-z) - \frac{(2 \mu(\vec r) + \lambda(\vec r) )}{2} \frac{t+2}{2} \frac{\bar u_{xx}^2 + \bar u_{zz}^2}{1 + \bar u_{xx}}  - \frac{\lambda(\vec r) }{2} \frac{t+2}{2} \frac{(\bar u_{zz} + \bar u_{xx})^2}{1 + \bar u_{zz}} \\
0 &= (2 \mu(\vec r) + \lambda(\vec r) )\bar u_{zz} +  \lambda(\vec r) \bar u_{xx} + \alpha_0(L_z-z) - \frac{(2 \mu(\vec r) + \lambda(\vec r) )}{2} \frac{t+2}{2} \frac{\bar u_{xx}^2 + \bar u_{zz}^2}{1 + \bar u_{zz}}  - \frac{\lambda(\vec r) }{2} \frac{t+2}{2} \frac{(\bar u_{zz} + \bar u_{xx})^2}{1 + \bar u_{xx}}. 
 \end{aligned}
\end{equation}
\end{widetext} 
Upon defining the following quantity,
\begin{eqnarray}
\gamma(z) \equiv \frac{\alpha_0 (L_z-z)}{2B_0} = \theta \frac{z-L_z}{L_x},
\end{eqnarray}
we can rewrite Eq.~(\ref{eq:fb_uyy}) as,
\begin{widetext} \normalsize 
\begin{equation}\label{eq:fb_uyy_2} 
\begin{aligned} 
0 &= \frac{2 \mu(\vec r) + \lambda(\vec r) }{B_0} \bar u_{xx} +  \frac{\lambda(\vec r)}{B_0} \bar u_{zz} + \gamma(z) - \frac{(2 \mu(\vec r) + \lambda(\vec r) )}{2B_0} \frac{t+2}{2} \frac{\bar u_{xx}^2 + \bar u_{zz}^2}{1 + \bar u_{xx}}  - \frac{\lambda(\vec r) }{2B_0} \frac{t+2}{2} \frac{(\bar u_{zz} + \bar u_{xx})^2}{1 + \bar u_{zz}} \\
0 &= \frac{2 \mu(\vec r) + \lambda(\vec r) }{B_0}\bar u_{zz} +   \frac{\lambda(\vec r)}{B_0}  \bar u_{xx} + \gamma(z) - \frac{(2 \mu(\vec r) + \lambda(\vec r) )}{2B_0} \frac{t+2}{2} \frac{\bar u_{xx}^2 + \bar u_{zz}^2}{1 + \bar u_{zz}}  - \frac{\lambda(\vec r) }{2B_0} \frac{t+2}{2} \frac{(\bar u_{zz} + \bar u_{xx})^2}{1 + \bar u_{xx}}. 
\end{aligned}
\end{equation}
\end{widetext}
When the perturbation is small, the strains $\bar u_{xx}$ and $\bar u_{zz}$ are on the order of $\gamma(z)$, 
\begin{eqnarray} \label{eq:req_small}
|\bar u_{zz}| = |\bar u_{xx} | \sim |\gamma(z)| \leq \left | \theta \frac{ L_z}{L_x} \right| \ll 1,
\end{eqnarray}
and we can eliminate the higher order quadratic terms $\sim \bar u_{ii}^2$ in Eq.~(\ref{eq:fb_uyy_2}) and retrieve the linear approximation displayed in Eqs.~(\ref{eq:uxx}):
\begin{eqnarray}
\bar u_{zz} = \bar u_{xx} =\frac{\alpha_0(z - L_z)}{2 B_0} + O(\gamma^2) 
\end{eqnarray}
The condition in Eq.~(\ref{eq:req_small}), for an approximately square sheet $L_x \sim L_z$, precisely corresponds to the small angle condition in Eq.~(\ref{eq:small_reg}). Thus, when the gravitational pressure is sufficiently weak, nonlinear effects can be neglected and the equilibrium deformation of particles with power law pair interactions is identical to that of the two-dimensional sheet with uniform elasticity constants. Since the Lennard-Jones potential is the sum of two power law terms, the previous statements also hold for the system studied in the main text, where the particles interact via the LJ potential in Eq.~(\ref{eq:LJ}). 

% In contrast, the space dependence of the Lam\'{e} coefficients has a first order effect on the localization phenomena, as shown by the linear dependence on $\theta$ in Eq.~\ref{eq:loc_fin}. 
% Henceforth, we will assume $\gamma \ll 1$ to be sufficiently small and use Eq.~\ref{eq:solx_lin} and \ref{eq:soly_lin} to approximate the equilibrium positions of the particles interacting with power-law pair potentials. 
\section{Phonon spectrum derivations}

\subsection{Free energy functional of the cone \label{sec:F0F}}

Upon letting $u_{ik} = \bar u_{ik} + \delta u_{ik}$, the free energy in Eq.~(\ref{eq:F_grav_cyl}) becomes,
\begin{eqnarray}
F&=&\frac{1}{2} \int d^{2} r\left[2 \mu \left( \bar u_{i k}^{2} + 2 \bar u_{ik} \delta u_{ik} + \delta u_{ik}^2 \right) \right. \\
&&\left. +\lambda \left( \bar u_{i i}^{2} + 2 \bar u_{jj} \delta u_{ii} + \delta u_{ii}^2 \right) +2 \alpha_{0}\left(L_{z}-z\right) (\bar u_{i i} + \delta u_{ii} ) \right]. \notag 
\end{eqnarray}
Upon grouping the terms linear and quadratic in $\delta u_{ik}$,
\begin{eqnarray}
F&=& F_0 + \frac{1}{2} \int d^{2} r 2 \delta u_{ik} \left[2 \mu \bar u_{ik} + \lambda \bar u_{jj} \delta_{ik} + \alpha_0(L_z - z) \delta_{ik} \right] \notag \\
&& + \frac{1}{2} \int d^{2} r \left[2 \mu \delta u_{ik}^2  +\lambda \delta u_{ii}^2  \right], 
\end{eqnarray}
where $F_0$ is the energy of the equilibrium configuration without displacements,
\begin{eqnarray}
F_0 =\frac{1}{2} \int d^{2} r\left(2 \mu \bar u_{i k}^{2}+\lambda \bar u_{i i}^{2}+2 \alpha_{0}\left(L_{z}-z\right) \bar u_{ii}\right),
\end{eqnarray}
we see that the coefficient of the term linear in $\delta u_{ik}$ vanishes according to Eq.~(\ref{eq:dFduii}).
% \begin{eqnarray}
% && 2 \mu \bar u_{ik} + \lambda \bar u_{jj} \delta_{ik} + \alpha_0(L_z - z) \delta_{ik} \\
% &= &2 \mu \bar u_{xx} + \lambda (\bar u_{zz} + \bar u_{xx}) + \alpha_0(L_z - z) = 0. 
% \end{eqnarray}
Hence, the change in free energy due to displacements away from equilibrium becomes,
\begin{eqnarray}
F = F_0 + \delta F = F_0 + \frac{1}{2} \int d^{2} r  \left[2 \mu \delta u_{ik}^2  +\lambda \delta u_{ii}^2  \right],
\end{eqnarray}
and $\delta F (\delta u_{ik})$ as a function of the new displacements takes the same form as that of the free energy of the unpressurized uniform 2d sheet. 
Thus, for a system with constant elastic coefficients $\mu$ and $\lambda$, the phonon spectrum of the deformed lattice would be the same as the undeformed lattice. However, for particles interacting via the Lennard-Jones potential, the inhomogeneous equilibrium configuration causes the elastic constants to become non-uniform $\mu, \lambda \rightarrow \mu(\vec r), \lambda (\vec r)$. As shown in Sec.~\ref{sec:phonons}, this has significant effects on the normal modes. 

% The equation of motion is given by
% \begin{eqnarray}
% \rho \frac{\partial^2  u_i (\vec x,t)}{ \partial t^2} = - \frac{\delta F}{\delta \vec u_i (\vec x, t)} = \left( \mu \delta_{ij} \partial_i \partial_i + (\mu + \lambda) \partial_i \partial_j \right) \delta u_j,
% \end{eqnarray}
% where $\rho$ is the mass density. Both the longitudinal mode (momentum parallel to displacement), 
% \begin{eqnarray}
% u_i = u_L \hat q_i e^{i \vec q \cdot \vec x - i \omega t}
% \end{eqnarray}and the transverse mode (momentum transverse to displacement)
% \begin{eqnarray}
% u_i = u_T (\hat x_i\times \hat q ) e^{i \vec q \cdot \vec x - i \omega t}
% \end{eqnarray}
% have phonon energies quadratic in $q$:
% \begin{eqnarray}
% m \omega_L^2 &=&  \left(2  \mu + \lambda\right) (q a)^2\\
% m \omega_T^2 &= & \mu (q a)^2. 
% \end{eqnarray}

\subsection{Results for power law interaction potentials \label{sec:power_law}}

In this section, we derive several key quantities for particles interacting via a power law potential of the following form,
\begin{eqnarray} \label{eq:pair_pot}
V_\text{int}(\vec x_n, \vec x_m) = \frac{A}{|\vec x_n - \vec x_m|^t},
\end{eqnarray}
where $A$ is some constant coefficient and $t$ is the exponent of the power law. Since the Lennard-Jones potential is the sum of two power law terms, the results are readily applied to the system in the main text. 

Note that when $t = 3$, Eq.~(\ref{eq:pair_pot}) captures the physics of experimentally relevant systems such as magnetic colloids, whose pair potential takes the form~\cite{gasser2010melting},
\begin{eqnarray} \label{eq:Ur}
U(r) = \frac{\mu_0 (\chi B)^2}{8 \pi} \frac{1}{r^3}
\end{eqnarray}
where $r$ is the distance between two colloids, $\chi$ is the magnetic susceptibility, $B$ is the magnitude of the external magnetic field, and $\mu_0$ is the permeability constant. However, note that particles experiencing a purely repulsive interaction potential such as Eq.~(\ref{eq:Ur}) requires an external potential to prevent them from flying apart. 
In contrast, since the Lennard-Jones potential has both a repulsive component $\sim 1/r^{12}$ and an attractive component $\sim 1/r^6$, a separate external potential is not needed to confine the particles. 

\subsubsection{Dispersion relation}

For interaction potentials of the form $V(r) = A/|r|^t$ ($t>0$), the matrix element in Eq.~(\ref{eq:H_dis_full}) becomes
\begin{eqnarray} 
\Pi_{ij}(\Delta \vec R) &=& A \left[ t ( t + 2) \frac{\Delta R_i \Delta R_j}{|\Delta \vec R|^{t+4}} - t\frac{\delta_{ij}}{|\Delta \vec R|^{t+2}} \right].\label{eq:pi_power}
\end{eqnarray}
To reduce notational clutter in the upcoming calculations, we abbreviate Eq.~(\ref{eq:pi_power}) as,
\begin{eqnarray}
\Pi_{ij}(\vec R) = \gamma \frac{R_i R_j}{|\vec R|^{t+4} } - \rho \frac{\delta_{ij}}{|\vec R|^{t+2}},
\end{eqnarray}
where,
\begin{eqnarray} \label{eq:gamma_rho}
\gamma \equiv A t(t+2), \quad \rho \equiv A t.
\end{eqnarray}

Upon following the procedure delineated in Sec.~\ref{sec:high_phonons_BE}, summing over the nearest neighbor lattice vectors $\{ \vec n_\alpha \}$, and using the following trigonometric identities,
\begin{eqnarray}
\cos(a + b) + \cos (a-b) &=& 2 \cos(a) \cos(b) \\
\cos(a + b) - \cos (a-b) &=& -2 \sin(a) \sin(b) \\
\cos (2a) &=& 2 \cos^2(a) - 1,
\end{eqnarray}
the dynamical matrix in momentum space $D_{ij}(\vec q)$ (see e.g. Eq.~(\ref{eq:Dij})) takes the following form, 
\begin{widetext} \normalsize 
\begin{equation} \label{eq:D_micro}
\begin{aligned}
D(\vec q) = \frac{4}{a^{t+2}} \begin{pmatrix} (\gamma - \rho)(1 - c_x^2) + (\frac{\gamma}{4} - \rho)( 1 - c_x c_z) & \gamma \frac{\sqrt{3}}{4} s_x s_z \\ \gamma \frac{\sqrt{3}}{4} s_x s_z & -\rho( 1 - c_x^2) + ( \frac{3 \gamma}{4} - \rho)(1 - c_x c_z) \end{pmatrix},
 \end{aligned}
\end{equation}
\end{widetext} 
where 
\begin{eqnarray}
c_x \equiv \cos \left(q_x \frac{a}{2}\right), &\quad& c_z \equiv \cos \left(q_z a \frac{\sqrt{3}}{2} \right) 
\end{eqnarray}
\begin{eqnarray}
s_x \equiv \sin \left(q_x \frac{a}{2} \right), &\quad& s_z \equiv \sin \left(q_z a \frac{\sqrt{3}}{2} \right). 
\end{eqnarray}

The eigenenergies near the Brillouin zone center (see Fig.~\ref{fig:LJ_disp}), e.g. along the $\Gamma$-$K$ line ($q_xa = qa \ll 1, ~q_z a = 0$) and along the $\Gamma$-$M$ line ($q_z a = qa \ll 1, ~q_x a = 0$), are given by
\begin{eqnarray}
\Omega_T &=& (qa)^2 \frac{3}{8} \frac{\gamma - 4 \rho}{a^{t+2}} = (qa)^2 \frac{3}{8} \frac{t^2 - 2 t}{a^{t+2}} \\
\Omega_L &=& (qa)^2 \frac{9}{8} \frac{\gamma - \frac{4}{3} \rho}{a^{t+2}} = (qa)^2 \frac{9}{8} \frac{t^2 - \frac{2}{3} t}{a^{t+2}}. 
\end{eqnarray}
Note that here, the longitudinal eigenenergy is not equal to 3 times the transverse eigenenergy. This change arises because the usual Cauchy condition~\cite{weiner2012statistical} is modified by the requirement of an external pressure to prevent the particles, which are interacting here via a purely repulsive potential, from flying apart. 

The band edge at point $M$ in the Brillouin zone on the longitudinal dispersion curve is given by
\begin{eqnarray}
\Omega_\text{edge}(A, t) &=& \frac{6}{a^{t+2}} ( \gamma - \frac{4}{3} \rho) = \frac{6A}{a^{t+2}} ( t^2 + \frac{2}{3} t). \label{eq:band_edge}
\end{eqnarray}
where we have used the values of $\gamma$ and $\rho$ from Eq.~(\ref{eq:gamma_rho}). It is now straightforward to calculate the band edge for the LJ potential in Eq.~(\ref{eq:LJ}) as
\begin{eqnarray}
\Omega_\text{edge}= \Omega(A = \varepsilon a_0^{12}, t=12) - \Omega(A = 2 \varepsilon a_0^6, t=6),
\end{eqnarray}
which gives Eq.~(\ref{eq:BE_LJ_NN}) in the main text.
% \begin{eqnarray} \label{eq:BE_LJ_NN_app}
% \Omega_\text{edge}^{} &=& \frac{6 \varepsilon}{a^{2}} \left[ \left( \frac{a_0}{a} \right)^{12} \left( 12^2 + \frac{2}{3} \cdot 12 \right) - 2 \left( \frac{a_0}{a} \right)^6 \left( 6^2 + \frac{2}{3} \cdot 6 \right) \right].
% \end{eqnarray}

\subsubsection{Effect of next nearest neighbors}

Because the Lennard-Jones potential falls off rapidly with distance, corrections to the spectrum due to further neighbors are expected to be small. Here, we examine the effects of incorporating next nearest neighbor (NNN) interactions on the spectrum, in particular the dispersion near the zone center and the band edge. 

\begin{figure}[htb]
    \centering
    \includegraphics[width = 0.75\columnwidth]{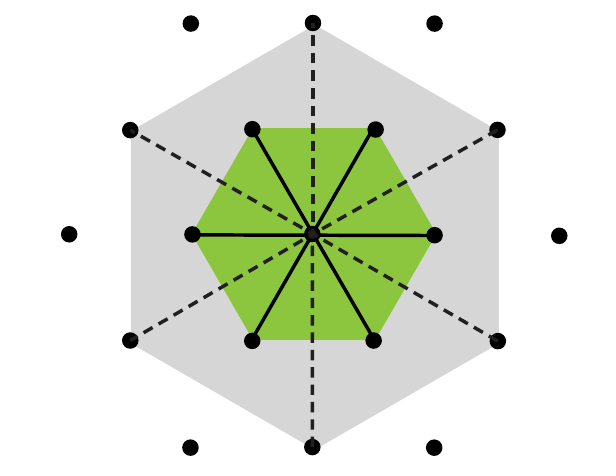}
    \caption{Schematic indicating the nearest neighbor (solid lines) and next nearest neighbor (dashed lines) interactions in real space on a triangular lattice.}
    \label{fig:NNN}
\end{figure}

For power law interactions of the form $V(r) = A/|r|^t$, the next nearest neighbor (see dashed lines in Fig.~\ref{fig:NNN}) contribution to the dynamical matrix is,
\begin{widetext} \normalsize 
\begin{equation} \label{eq:D_micro_NNN}
\begin{aligned}
D^{NNN}(\vec q) = \frac{4}{(\sqrt{3} a)^{t+2}} \begin{pmatrix} - \rho(1 - \bar c_z^2) + (\frac{3 \gamma}{4} - \rho)( 1 - \bar c_x \bar c_z) & \gamma \frac{\sqrt{3}}{4} \bar s_x \bar s_z \\ \gamma \frac{\sqrt{3}}{4} \bar s_x \bar s_z & (\gamma -\rho)( 1 - \bar c_z^2) + ( \frac{ \gamma}{4} - \rho)(1 - \bar c_x \bar c_z) \end{pmatrix},
 \end{aligned}
\end{equation}
\end{widetext} 
where,
\begin{eqnarray}
\bar c_x \equiv \cos(3 q_x a/2), &\quad& \bar c_z \equiv \cos(q_z a \sqrt{3}/2), \\
\bar s_x \equiv \sin(3 q_x a/2), &\quad& \bar s_z \equiv \sin(q_z a \sqrt{3}/2). 
\end{eqnarray}
Near the zone center (e.g. $q_z = 0,~ q_x = q \approx 0$), the contribution to the dynamical matrix from next nearest neighbors is equal to the nearest neighbor contribution scaled by an overall constant,
\begin{eqnarray}
D^{NNN}(q \approx 0) = \frac{1}{3^{t/2}} D^{NN}(q \approx 0).
\end{eqnarray}
Thus, the addition of next nearest neighbor interactions scales the slopes of both dispersion branches near the zone center by the same constant factor and doesn't alter the Poisson ratio. 

The shift in band edge due to NNN interactions is,
\begin{eqnarray}
\Delta \Omega_\text{edge}^{NNN} (A,t) &=& \frac{2 A}{\left( \sqrt{3}a\right)^{t+2} }  \left( t^2 - 2  t \right) .
\end{eqnarray}
For a uniform lattice ($a = a_0$) of particles with LJ interactions (Eq.~(\ref{eq:LJ})), the NNN modification becomes,
\begin{eqnarray}
\Delta \Omega_\text{edge}^{NNN} &=& - \frac{784}{729} \frac{\varepsilon}{a_0^2} \approx - 1.0754 \frac{\varepsilon}{a_0^2},
\end{eqnarray}
which is indeed very small compared to the nearest neighbor contribution in Eq.~(\ref{eq:BE_LJ}). We thus neglect the effect of next nearest neighbors in our calculations.

\subsection{Real space dynamical matrix $\bf{M}$ \label{app:M_calc}}

The eigenvalue problem in Eq.~(\ref{eq:eig_prob}) can be solved by diagonalizing the dynamical matrix $\bf{M}$ given by,
\begin{eqnarray}
{\mathbf{M}} = 
\begin{bmatrix}
\bf \tilde \Pi_{11} & \bf \tilde \Pi_{12} \\ \bf \tilde \Pi_{21}& \bf \tilde \Pi_{22} 
\end{bmatrix}.
\end{eqnarray}
For fixed displacement boundary conditions at the bottom, each $\bf \tilde \Pi_{ij}$ is a block matrix of size $(N-N_\text{row}) \times (N-N_\text{row})$, where $N_\text{row} \equiv L_x / a_0$  is the number of particles in one horizontal row (the triangular lattice is oriented such that the top and bottom edges are flat). Diagonalizing $\bf M$ then gives us a total of $2(N - N_\text{row})$ normal modes. The off-diagonal elements $\tilde \Pi_{ij}^{(n \neq m)}$are given by
\begin{eqnarray}
{\bf \tilde \Pi_{ij}}^{(nm)} = \Pi_{ij}(\vec R_n - \vec R_m)
\end{eqnarray}
where $\Pi_{ij}(\vec R)$ is given in Eq.~(\ref{eq:Pi_ij}), $\vec R_n$ is the equilibrium position of the $n$-th particle, and $n$ and $m$ only include the $N- N_\text{row}$ particle indices that exclude those in the bottom fixed row. 
The diagonal elements are given by,
\begin{eqnarray}
{\bf \tilde \Pi_{ij}}^{(nn)} = -\sum_{m'\neq n}\Pi_{ij}(\vec R_n - \vec R_{m'})
\end{eqnarray}
where $m'$ is summed over all $N$ particle indices, including those in the bottom row. Finally, the eigenvector ${\bf u}^{(\alpha)}$ takes the form,
\begin{eqnarray}
{{\bf u}^T}^{(\alpha)} &=& \left[ u_1^{(\alpha)} (\vec R_1), \cdots, u_1^{(\alpha)} (R_{N-N_\text{row}}), \right. \\
&&\left. u_2^{(\alpha)} (\vec R_1), \cdots, u_2^{(\alpha)} (R_{N-N_\text{row}}) \right]. \notag
\end{eqnarray}

\bibliography{references.bib}
\end{document}